\documentclass[aps,prd,twocolumn,amsmath,amssymb,nofootinbib,10pt]{revtex4-1}

\pdfoutput=1

\usepackage{amsfonts}
\usepackage{amsmath}
\usepackage{amssymb}
\usepackage{amsthm}
\usepackage{graphicx,xcolor}
\usepackage[footnotesize]{caption}
\usepackage{slashed}
\usepackage{hyperref}
\usepackage{multirow}
\usepackage{subcaption}
\usepackage{dsfont}
\usepackage{mathrsfs}
\usepackage[utf8]{inputenc}
\usepackage{placeins}
\graphicspath{{./figures/}}

\DeclareMathAlphabet{\boldmathe}{T1}{cmr}{bx}{it}
\newcommand{\mbf}[1]{\boldmathe{#1}}
\newcommand{\Nf}{N_{\mathrm{f}}}
\newcommand{\Ns}{N_{\mathrm{s}}}
\newcommand{\Nt}{N_{\mathrm{t}}}
\newcommand{\kmax}{k_{\mathrm{max}}}
\newcommand{\nmax}{\nu_{\mathrm{max}}}
\newcommand{\Z}{\mathbb{Z}}
\newcommand{\Tc}{T_{\mathrm{c}}}
\renewcommand{\L}{\mathcal{L}}
\newcommand{\ii}{\mathrm{i}}
\newcommand{\Seff}{S_{\mathrm{eff}}}

\newcommand{\argmax}{\mathrm{argmax}}

\newcommand{\vx}{\mbf{x}}
\newcommand{\vy}{\mbf{y}}

\newcommand{\sref}[1]{Sec.~\ref{#1}}
\newcommand{\aref}[1]{App.~\ref{#1}}
\renewcommand{\eqref}[1]{Eq.~(\ref{#1})}
\newcommand{\tref}[1]{Tab.~\ref{#1}}
\newcommand{\fref}[1]{Fig.~\ref{#1}}

\begin{document}

	\title{Inhomogeneities in the $2$-Flavor Chiral Gross-Neveu Model}
	
	\author{Julian J. Lenz}
	\email{julian.johannes.lenz@uni-jena.de}
	\author{Michael Mandl}
	\email{michael.mandl@uni-jena.de}	
	\author{Andreas Wipf}
	\email{wipf@tpi.uni-jena.de}		
	
	\affiliation{Theoretisch-Physikalisches Institut, Friedrich-Schiller-Universität Jena,  D-07743 Jena, Germany}
	
	\date{\today}
	
\begin{abstract}	
We investigate the finite-temperature and -density
chiral Gross-Neveu model with an axial U$_A$(1) symmetry
in $1+1$ dimensions on the lattice. In the limit where the number of
flavors $\Nf$ tends to infinity the continuum model has been solved 
analytically and shows two phases: a symmetric 
high-temperature phase with a vanishing condensate and a low-temperature
phase in which the complex condensate forms a chiral spiral which
breaks translation invariance.
In the lattice simulations we employ
chiral SLAC fermions with exact axial symmetry.
Similarly to $\Nf\to\infty$, we find for $8$ flavors,
where quantum and thermal fluctuations are suppressed, two distinct regimes in the 
$(T,\mu)$ phase diagram, characterized by qualitatively different behavior 
of the two-point functions of the condensate fields. More surprisingly, 
at $\Nf=2$, where fluctuations are no longer suppressed, the model still
behaves similarly to the $\Nf\to\infty$ model and we conclude that the chiral 
spiral leaves its footprints even on systems with a small number of flavors. 
For example, at low temperature 
the two-point functions are still dominated by chiral spirals with pitches proportional to the inverse chemical potential,
although in contrast to large-$\Nf$ their amplitudes decrease with distance. We argue that these results should not be interpreted as the spontaneous breaking of a continuous symmetry, which is forbidden in two dimensions. Finally, using Dyson-Schwinger
equations we calculate the decay
of the U$_A$(1)-invariant fermion four-point function
in search for a BKT phase at zero temperature.
\end{abstract}		
	\maketitle

	\section{Introduction}
	\label{sec:introduction}
	A surprising amount of physical phenomena in particle-
and condensed-matter physics are well described by 
four-Fermi theories. For instance, they are employed to model
low-energy chiral properties of Quantum Chromodynamics (QCD).
The effective four-Fermi theory describing the dynamics of nucleons 
and mesons goes back to Nambu and Jona-Lasinio (NJL) \cite{NJL61}
and is built upon interacting Dirac fermions with chiral symmetry, 
paralleling the construction of Cooper pairs from electrons in the BCS 
theory of superconductivity.  

In fact, most of our knowledge about 
QCD at intermediate baryon densities stems from the study of 
NJL-type effective theories, since in this
regime one needs non-perturbative methods but cannot use 
lattice 
field theory techniques due to the complex-action problem. 
In a similar spirit, a four-Fermi current-current interaction 
among leptons (and quarks) was proven to give an accurate
phenomenological description of the weak interaction at low 
energy $p^2\ll m_\mathrm{W}^2$. In the pioneering work by 
E. Fermi the currents are made up from the proton, neutron,
electron and neutrino fields \cite{Fermi:1934hr}.
In four spacetime dimensions interacting Fermi theories, such as the 
NJL model or Fermi theory, are non-renormalizable and 
thus can only serve as effective (low-energy) approximations
which need to be UV completed. For the two
examples given these completions are of course known.

The dynamical creation of a condensate from strong fermion interactions
as seen in NJL-type models inspired many theories of the breaking 
of electroweak symmetry, such as technicolor (see the review
\cite{Hill:2002ap}) and the top-quark condensate
\cite{Miransky:1988xi}.

Four-Fermi theories in two spacetime dimensions are renormalizable
and asymptotically free (some are integrable or even soluble)
and share certain features with their cousins in four dimensions.
The most prominent examples are the Thirring model with a
current-current interaction \cite{Thirring:1958in},
which is $S$-dual to the sine-Gordon model, and the Gross-Neveu (GN)
model with a scalar-scalar interaction  \cite{GN74}, which serves as a
toy model for the theory of strong interactions.

With the discovery of novel materials (like Dirac and Weyl semimetals
in two and three spatial dimensions) and the development of 
experimental techniques (for example optical lattices to trap atoms)
we have witnessed a steadily increasing interest in models describing 
interacting fermions. Such models in lower dimensions
describe one-dimensional and planar systems, such as polymers
\cite{CB82,CM94,Cal11,Kun19,TU05}, graphene \cite{EB19,EKK19} or 
high-$T_c$ superconductors \cite{Liu99,ZKK01}, to name some prominent 
examples.

Interacting Fermi theories at finite temperature and density
were mainly investigated in the limit of a large number of fermion 
flavors $\Nf$. For $\Nf\to\infty$ the saddle-point approximation 
becomes exact and one can solve the corresponding gap equation 
analytically on the set of homogeneous condensates. But,
for the $(1+1)$-dimensional GN model at low temperature 
and large chemical potential the relevant solutions of the gap
equation are actually inhomogeneous in space. They have been
constructed in \cite{Thi04} for the GN model with discrete 
and in \cite{ST00,ST01r} for the chiral GN model with continuous chiral 
symmetry. These remarkable analytic results for $\Nf\to\infty$ prove the existence of 
inhomogeneous phases,
which are regions in parameter space where the chiral condensate acquires 
a spatial dependence, indicating the spontaneous breakdown of not only 
chiral symmetry alone but in a combination with spacetime symmetries
(see \cite{Thi06} for a review).

Are these inhomogeneous phases at large densities 
an artifact of the large-$\Nf$ limit as suggested by various
no-go theorems in two spacetime dimensions? 
To address this question, a better understanding of
interacting Fermi systems at finite $\Nf$ with
regards to inhomogeneous phases is required.
But the spontaneous breaking of translation invariance
is not merely of academic interest: Systems where an inhomogeneous
state develops spontaneously have been extensively discussed in the
condensed-matter literature. A prominent example is the inhomogeneous
pairing inside a superconductor in a magnetic field, 
predicted by Larkin, Ovchinnikov, Fulde and Ferrell (LOFF phase) \cite{FF64,LO64}. Similar
types of pairings can occur in many other physical
systems, ranging from supersolids to ultracold atomic gases (see the 
reviews \cite{Casalbuoni:2003wh,Bloch:2008zzb}). 
The UV cutoff which is inherent in all condensed-matter systems
inhibits a direct translation of these findings to 
quantum field theory and particle physics where one removes the 
cutoff during the process of renormalization. 

A first attempt to investigate the fate of inhomogeneous phases at finite $\Nf$ has been 
made in recent lattice studies \cite{LPW20,LPW20_1}, where 
the existence of spatially varying chiral condensates in the
$(1+1)$-dimensional GN model with $2$, $8$, and $16$ flavors was confirmed. 
The present work serves as a follow-up, 
providing a similar analysis of the \emph{chiral Gross-Neveu} (cGN) 
model with a continuous axial symmetry, characterized by the Lagrangian
\begin{equation}\label{eq:cgn_lagrangian}
	\L = \bar{\psi}\ii\slashed{\partial}\psi +\frac{g^2}{2\Nf}\left((\bar{\psi}\psi)^2+(\bar{\psi}\ii\gamma_*\psi)^2\right)\;,
\end{equation}
where $g^2$ denotes a dimensionless coupling constant and the
two-dimensional matrix $\gamma_*=i\gamma_0\gamma_1$ 
is the analog of $\gamma_5$ in two spacetime dimensions. The summation 
over $\Nf$ flavors of fermions is implied in the fermion bilinears entering
\eqref{eq:cgn_lagrangian}. 

Below we shall see that the results of our simulations with chiral SLAC fermions 
resemble the analytical findings of the large-$\Nf$ limit \cite{ST00, ST01r}. 
The analysis of the GN model in \cite{LPW20} has already given 
clear evidence that the chiral and doubler-free SLAC fermions and 
naive fermions yield comparable results in the continuum limit, with the 
former converging considerably faster.\footnote{The same observation applies
to supersymmetric Yukawa models \cite{BKU08,Kastner:2008zc}.} Using
SLAC fermions has the additional advantages that the lattice cGN model
is invariant under axial U$_A$(1) transformations and that we can study
the system with $\Nf=2$ without encountering a sign problem. With naive
fermions the GN and cGN models have no sign problems only for $\Nf$ a multiple of
$8$. In the present work, however, we want to investigate how much the models at
finite flavor number differ from the analytic solutions at infinite $\Nf$, 
for which $\Nf=8$ might be too large, see \cite{LPW20}. We do not use Wilson fermions since we are mainly 
interested in the chiral properties of cGN models. Staggered fermions,
on the other hand, may lead to wrong results for interacting Fermi systems, 
as has been demonstrated in \cite{Wellegehausen:2017goy,Lenz:2019qwu,Hands:2020itv}.

Our work is organized as follows. In Sec. \ref{sec:analytical} 
we summarize relevant facts about the finite-temperature and -density
cGN model with Lagrangian (\ref{eq:cgn_lagrangian})
in the continuum, which will be used in the subsequent sections.
In Sec. \ref{sec:lattice} the lattice cGN model with chiral SLAC fermions
is presented, relevant observables are introduced and the lattice setup is
discussed. Section \ref{sec:results} contains our simulation results on the 
inhomogeneous condensation of the scalar and pseudo-scalar bilinears and 
their interrelation.  We calculate the phase diagram in the $(T, \mu)$ plane
for various lattice sizes and lattice constants in order to study the
thermodynamic and continuum limits.
We shall see that even for the smallest accessible value $\Nf=2$ the results
resemble those for the exact solution of the system with
$\Nf\to\infty$. Towards the end we exploit Dyson-Schwinger equations to
study the U$_A(1)$-invariant fermion four-point function in the infrared.

	\section{Analytical considerations}
	\label{sec:analytical}
	\subsection{Symmetries and reformulations}
The chiral GN model with Lagrangian (\ref{eq:cgn_lagrangian}) most 
prominently features a global axial U$_A$(1) symmetry,
\begin{equation}\label{eq:chiral_symmetry}
	\psi(\vx) \rightarrow e^{\ii\alpha\gamma_*}\psi(\vx)\;, \quad \bar{\psi}(\vx) \rightarrow \bar{\psi}(\vx)\,e^{\ii\alpha\gamma_*}\;,
\end{equation}
with a continuous parameter $\alpha\in\mathbb{R}$.
In this work we denote spacetime coordinates
by bold letters, for example
\begin{equation}
\vx=\begin{pmatrix}t\\ x\end{pmatrix}\,.
\end{equation}
The continuous axial symmetry is to be compared with the 
discrete $\Z_{2}$ symmetry of the model considered in \cite{LPW20,LPW20_1}. 
Further symmetries of the model include a flavor-vector symmetry that ensures 
the factorization of the fermion determinant, parity and charge 
conjugation symmetry responsible for the absence of the sign problem
for even $\Nf$ (see \cite{LPW20} for details) and, of course, Euclidean 
spacetime symmetry. 

As is usually done we introduce the complex auxiliary field
$\Delta$ in order to bring the 
Lagrangian (\ref{eq:cgn_lagrangian}) to the equivalent form
\begin{equation}\label{eq:lagrangian_no_mu}
	\L = \ii\bar{\psi}\left(\slashed{\partial}+P_{+}\Delta + P_{-}\Delta^{*}\right)\psi + \frac{\Nf}{2g^2}|\Delta|^{2}\;,
\end{equation}
where $P_{\pm} = \frac{1}{2}\left(\mathds{1} \pm \gamma_*\right)$ are the 
chiral projectors. This Lagrangian is invariant under the axial transformations
(\ref{eq:chiral_symmetry}) supplemented by 
\begin{equation}
\Delta(\vx)\mapsto e^{-2\ii\alpha}\Delta(\vx)\,.\label{eq:chiral_symmetryb}
\end{equation}
One can show the equivalence of Lagrangians 
(\ref{eq:lagrangian_no_mu}) and (\ref{eq:cgn_lagrangian}) by using
the equations of motion for the auxiliary field $\Delta$. This 
equivalence persists on the quantum level 
because the $\Delta$ integration in the path integral is Gaussian and can 
be done analytically, leading back to \eqref{eq:cgn_lagrangian}. It is no 
more difficult to obtain the following Dyson-Schwinger (DS)
equations relating the expectation values of the auxiliary fields to 
the symmetry-breaking chiral condensates:\footnote{In $1+1$ dimensions
	the condensates vanish for finite $\Nf$. Later we shall study
	DS equations for bilinears of the condensate fields.}
\begin{equation}\label{eq:ward_identities}
	\langle\bar{\psi}P_{+}\psi\rangle = \frac{\ii\Nf}{2g^2}\langle\Delta^{*}\rangle\;,\quad  \langle\bar{\psi}P_{-}\psi\rangle = \frac{\ii\Nf}{2g^2}\langle\Delta\rangle\;.
\end{equation}
For later use we introduce two further parametrizations of $\Delta$ in terms 
of its real and imaginary parts $\sigma$ and $\pi$ and in terms of its absolute value $\rho$ and 
phase $\theta$:
\begin{equation}\label{eq:delta}
	\Delta = \sigma + \ii\pi = \rho\, e^{\ii\theta}\;.
\end{equation}
In order to study finite baryon densities we also introduce a chemical 
potential $\mu$ for the fermion number density $\bar{\psi}\gamma_0\psi$, such 
that the Lagrangian takes the form
\begin{equation}\label{eq:lagrangian}
	\L = \bar{\psi}\ii \mathcal{D}\psi + \frac{\Nf}{2g^2}\rho^2\;,
\end{equation}
where the Dirac operator $\mathcal{D}$ is defined as
\begin{equation}\label{eq:dirac_operator}
	\mathcal{D} = \slashed{\partial} + \mu\gamma_0 + \rho e^{\ii\gamma_*\theta}\;.
\end{equation}
It is understood that this operator acts on all flavors in the same way, such that in 
the multi-flavor case we may use the same symbol as for one flavor.

While there is no gauge invariance in this model, one can still trade 
the compact field  $\theta$ for an imaginary vector potential
\begin{align}
A_{\mu} = \frac{\ii}{2}\varepsilon_{\mu\nu}\partial_{\nu}\theta \qquad
(\varepsilon_{01}=1)
\end{align}
in the following sense:
\begin{align}
\mathcal{D} = e^{\ii\gamma_{*}\theta/2}\left(\ii\slashed{D} + \rho\right)e^{\ii \gamma_{*}\theta/2}\;,
\end{align}
where the covariant derivative $D_{\mu}$ is defined as
\begin{equation}
D_{\mu} = \partial_{\mu} -\ii A_{\mu} + \mu\delta_{\mu 0}\;.
\end{equation}
Since the main focus in our work is on homogeneous and
inhomogeneous phases of the finite-temperature and finite-density cGN model we impose that $\psi$, $\bar{\psi}$ 
are anti-periodic and $\Delta$, $\Delta^*$ are periodic in Euclidean
time with period $\beta$, where $\beta$ is the inverse temperature.
We furthermore impose that all fields are periodic in the spatial direction
with period $L$.

Integrating out the fermions in the partition function yields an effective 
bosonic theory in which the auxiliary bosons become dynamical via fermion loops,
\begin{equation}
	\label{eq:partition_function}
	Z = \int\!\mathscr{D}\Delta\; e^{-\Nf\Seff[\Delta]}\;,
\end{equation}
with the effective action
\begin{equation}
	\label{eq:seff}
	\Seff[\Delta] = -\ln \det \mathcal{D} +\frac{1}{2g^{2}}\int d^2x\,\rho^{2}(\vx)\;.
\end{equation}
We used that the fermion determinant of the multi-flavor model
is $(\det \mathcal{D})^{\Nf}$ 
with the one-flavor operator  $\mathcal{D}$ appearing in \eqref{eq:seff}.
A convenient (and widely adopted) way of renormalizing this formal expression is 
a choice of the bare coupling $g^{2}$ such that $\Seff$ for $T=0$ and $\mu=0$ takes 
its global minimum at some prescribed positive value $\rho(t,x)=\rho_{0}$. 
The corresponding gap equation in the thermodynamic limit,
\begin{equation}
	\frac{1}{g^{2}} = \frac{1}{2\pi}\int_0^\Lambda\frac{p\,d p}{p^{2} + \rho_{0}^{2}}\;,
\end{equation}
yields the cutoff dependence of the bare coupling.
\subsection{Large-\texorpdfstring{$\Nf$}{Nf} results}
In the large-$\Nf$ limit the saddle-point approximation to
the path integral (\ref{eq:partition_function}) becomes exact and the 
\emph{grand potential} $\Omega$ proportional to the minimum of the
effective action (\ref{eq:seff}) on the space of auxiliary fields,
\begin{equation}
L\Omega(T,\mu,L)=-\frac{1}{\Nf}\log Z\stackrel{\Nf\to\infty}{\longrightarrow}
\min_{\Delta}\, \Seff[\Delta]\;.
\end{equation}
This means that in the large-$\Nf$ limit the path integral is localized
at the minimizing configuration $\Delta_\mathrm{min}$. It follows, for example, 
that the expectation value of $\Delta$ is equal to $\Delta_\mathrm{min}$.

The condition of a (local) minimum, maximum or saddle point is 
expressed by the gap equation
\begin{align}
	0 = \frac{\delta\Seff }{\delta\Delta^{*}}\;,
\end{align}
which has been extensively studied in the literature. 
A constant solution $\Delta$ of this equation can be mapped
into the constant real solution $\vert\Delta\vert$ by
an axial rotation. But, for real $\Delta$ the effective
action of the cGN model simplifies to that of the
GN model. Hence, if $\rho_0$ solves
the GN gap equation then $\rho_0e^{\ii\theta}$ with constant $\theta$
solves the cGN gap equation.

\begin{figure}[h!]
	\includegraphics[width=0.95\linewidth]{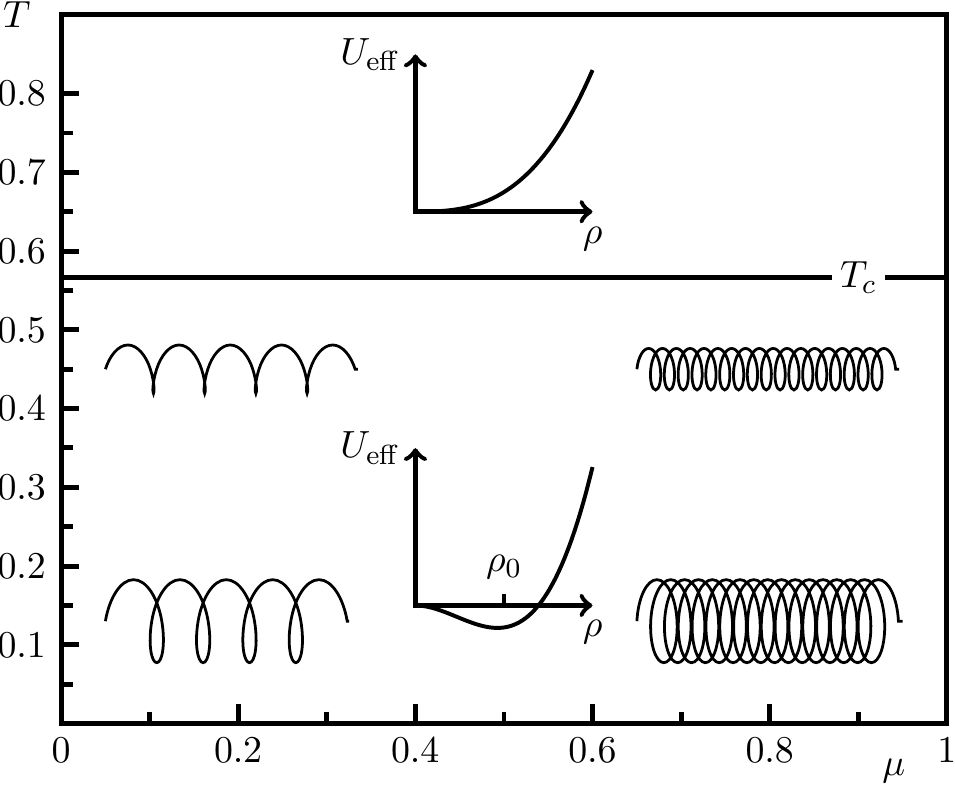}
	\caption{
		The phase diagram of the cGN model in the large-$\Nf$
		limit (see \cite{ST01r}). One critical temperature $\Tc$ for all 
		$\mu$ marks the transition from chiral spirals 
		at $T<\Tc$ to the symmetric phase at $T>\Tc$.
	Units are set by the condensate $\rho_0$ at zero temperature and
	zero chemical potential.}
	\label{f:large_n_phase_diagram}
\end{figure}

On can show that for temperatures above the critical temperature
\begin{equation}\label{eq:large_N_critical_temperature}
\frac{\Tc}{\rho_{0}} = \frac{e^{\gamma}}{\pi}\approx 0.567
\end{equation}
and for all $\mu$ the cGN model (in the large-$\Nf$ limit) is in a symmetric phase with a vanishing 
condensate field \cite{BDT09}.
Here $\gamma$ is the Euler-Mascheroni constant.
More surprising is the fact that below $T_c$ and for all $\mu\neq 0$ 
there are no homogeneous solutions of the gap equation
which minimize $\Seff$. Instead, the minimizing configurations
are helixes with pitch $\pi/\mu$,
\begin{align}
	\label{eq:chiral_spiral}
	\Delta(\vx) =\Delta(x)= \bar{\rho}(T)\, e^{2\ii k(\mu) x}\;,
\end{align}
so-called chiral spirals, with a temperature-dependent amplitude $\bar{\rho}(T)$ and $k(\mu)=-\mu$ in the large-$\Nf$ limit. For vanishing
chemical potential the chiral spiral degenerates to a homogeneous 
configuration, which relates to the large-$\Nf$ solution of the $\Z_2$ GN model
at $
\mu=0$. We conclude that the profile function 
$\bar{\rho}(T)$ is just the condensate of the GN model at $\mu=0$, 
which decreases monotonically in $T$ until it vanishes at $T_c$. 
The large-$\Nf$ phase diagram in the $(T, \mu)$ plane is depicted in 
\fref{f:large_n_phase_diagram}.

\subsection{Spontaneous symmetry breaking in low dimensions}
Under rather natural assumptions the
existence of \emph{perfect} long-range order (as opposed to \emph{quasi}-long-range order) 
in lower dimensions is excluded by the 
celebrated Coleman-Hohenberg-Mermin-Wagner theorem
\cite{LL1958,MW66,Hoh67,Col73}.
This theorem states that continuous symmetries cannot be 
spontaneously broken at finite $T$ in low-dimensional systems with
short-range interactions. 
In particular, for zero-temperature systems the theorem says:
\emph{The continuous symmetries cannot be spontaneously broken in $(1+1)$-dimensional quantum  systems.}
If a continuous symmetry were spontaneously broken, then the system 
would contain Goldstone bosons, which is impossible in two
spacetime dimensions because massless scalar fields have an IR-divergent behavior \cite{Col73}. Discrete symmetries, on the other hand, can still be spontaneously broken in two
dimensions.

There is a domain-wall proof of the theorem, of which the basic intuition 
is to rotate the field or, in a spin-model language, the values
of spins in a finite region with an arbitrarily small energy cost. 
This is achieved by creating a domain wall of finite thickness 
interpolating between the regions with rotated and 
unrotated spins. If the symmetry group were discrete, there would be
no smooth interpolation and hence a finite cost for creating 
domain walls.

The increasing strength of fluctuations (thermal and quantum)
in the IR with decreasing dimension $d$ is known
from the (Euclidean) free scalar field with propagators
\begin{equation}
\langle\phi(\vx)\phi(0)\rangle
\stackrel{m\to 0}{=} 
\begin{cases}
-\frac{1}{2}\vert x\vert&d=1\;,\\
-\frac{1}{2\pi}\log\vert \vx\vert&d=2\;,\\
\frac{1}{4\pi\vert \vx\vert}&d=3\;.
\end{cases}
\end{equation}
The interpretation of the IR divergence in $d=1$ and $2$
is that the field fluctuations cannot stay centered around a 
mean.
It implies that far away from a given spacetime point
the field takes completely different values than at the given point.
This happens in one and two dimensions where the fluctuations
move the field arbitrarily far from an initial value such that
it has no well-defined average. 

This reasoning should apply to translation invariance as well:
If the distance between two neighboring particles
on a \emph{wire} fluctuates by $\delta x$, then the $n$th particle's separation 
fluctuates as $\sqrt{n}\,\delta x$ and thus
diverges for large $n$.  These large fluctuations destroy any
long-range order in the position of the particles
and R. Peierls concluded that a one-dimensional 
equally spaced chain with one electron per ion is unstable \cite{Pei34}.
In higher dimensions ($d\geq 3$) the fluctuation-induced 
correlations fall off at large distances and are not
strong enough to destroy long-range order.

Furthermore, based on the powerful energy-entropy argument 
it has been argued that any spontaneous symmetry breaking (SSB) should be disallowed in $1+1$ dimensions
at finite temperature \cite{LL1958}. In the argument one
considers a small number $N$ of local perturbations of an ordered
state (e.g. aligned spins in the Ising model). The entropic contribution of
these perturbations to the free energy is $\propto N\ln N$ while 
the energy penalty is only $\propto N$. Thus, the entropic contribution 
can overcome the energy barrier and destroy the order. This perspective 
is directly applied to the discrete GN model in \cite{DMR75}.

Hence, the breaking of translation invariance in the $(1+1)$-dimensional GN model seems to be excluded on general grounds. On the other hand, the no-go theorems do not apply in the large-$\Nf$ limit where the analytical solution shows that the finite-temperature
and finite-density equilibrium state is not translation invariant.
What may happen at finite $\Nf$ is a subtle issue and has
been discussed (including the underlying assumptions 
of certain no-go theorems) in \cite{LPW20}.

Besides the questions raised in \cite{LPW20} there are more points
to be clarified with regards to the applicability of the no-go theorems:	
It is not obvious whether the effective action
$\Seff[\Delta]$ containing the non-local fermion determinant 
is short ranged enough to ensure the convergence of certain integrals,
which is assumed in \cite{MW66}. 
Although \cite{Hoh67} treats fermions as well, the result is based on
sufficient convergence (in form of $f$ sum rules) and gives 
itself an example of violation.

We emphasize that the no-go theorems allow for a BKT phase with
quasi-long-range order expressed by slowly decaying correlations
$\propto 1/\vert \vx\vert^{\alpha}$ and a BKT transition to 
a massive phase
with short-range correlations $\propto e^{-m\vert \vx\vert}$ 
\cite{Berezinsky:1970fr,Kosterlitz:1973xp}.
There is no symmetry breaking and no order parameter
involved in the strict sense, but the slowly decaying 
correlations of a BKT phase allow for large regions 
of one distinguished local state.

\subsection{Perturbations of chiral spirals}
\label{sec:perturbations}
How are the inhomogeneous phases of the GN and cGN models in the
large-$\Nf$ limit compatible with the no-go theorems discussed above?
In a way the large parameter $\Nf$ takes over 
the role of an extra spatial dimension. For example, in the 
domain-wall argument given above the energy penalty is
multiplied by the large number $\Nf$ and in the limit $\Nf\to\infty$
may overcome the entropy 
gain.

This and further heuristic arguments can be substantiated by a
systematic expansion in the small parameter $1/\Nf$, whereby one
assumes that for finite $\Nf$ the continuous U$_A$(1) axial symmetry is not spontaneously 
broken. Under an axial rotation the radial 
field $\rho$ is left invariant and $\theta$ is shifted by
a constant. This means that an invariant effective action is a functional
of the form \cite{Witten:1978qu}
\begin{equation}
\Seff=\Seff[\rho,\partial_\mu\theta]\;.
\end{equation}
This effective action is used to calculate expectation
values of functions of $\Delta=\rho e^{\ii\theta}$ and its 
complex conjugate field $\Delta^*$. However, in the continuum model a
condensate $\langle\Delta\rangle$ cannot form (it would break
the axial symmetry) and with
chiral SLAC fermions and the ergodic
rHMC algorithm it averages out in lattice simulations, see
Sec. \ref{sec:setup}.
Thus, following our previous studies \cite{LPW20,LPW20_1}, the correlator
\begin{equation}\label{eq:deltastar_delta_correlator}
C(x) = \left\langle \Delta^{*}(t,x)\Delta(t,0) \right\rangle 
\end{equation}
will be of particular interest to us. 

For $\Nf\to\infty$ the path integral is localized at the 
chiral spiral (\ref{eq:chiral_spiral}) and we find
\begin{equation}
C(x) = \bar{\rho}^{2}e^{-2\ii kx}\;.\label{eq:Cspiral}
\end{equation}
Clearly, for finite $\Nf$ we must admit small deviations from the chiral spiral,
\begin{equation}
\Delta(\vx)=\big(\bar\rho+\delta\rho(\vx)\big)\, e^{2\ii kx+\ii \delta\theta(\vx)}\;,
\end{equation}
and expand the effective action in powers of the fluctuation fields $\delta\rho$ 
and $\delta\theta$. An explicit calculation at zero temperature and in an infinite volume shows that the term linear in the fluctuation fields vanishes if the bare GN coupling 
depends on $\bar\rho$
according to
\begin{equation}
\frac{1}{g^2}=
\frac{1}{2\pi}
\log\frac{\Lambda^2+\bar{\rho}^2}{\bar{\rho}^2}\quad\text{and}\quad
k+\mu=0\;.\label{gap3}
\end{equation}
The first relation is recognized as the gap equation of the $\Z_2$ GN model. 
For large volumes the wave number
$k$ becomes continuous and the second relation
can be fulfilled for all $\mu$. Since the effective
action only depends on $k$ via $k+\mu$, 
this relation implies that
$S_\mathrm{eff}$ is independent of both $k$ and $\mu$.
In a finite box with quantized $k$, however, 
the background field $\bar\rho$ and
the effective action will generically
depend on $k+\mu$.

The contribution quadratic in the fluctuation fields is rather 
lengthy and has divergent terms which
all cancel when one uses the gap equation (\ref{gap3}).
If in addition the wave number of the chiral spiral
obeys $k+\mu=0$, then one obtains
\begin{equation}
\begin{aligned}
&S_\mathrm{eff}=
V U_\mathrm{eff}(\bar{\rho})
+\frac{1}{2\pi}\int \,
\delta\rho\,K_\varDelta\,
\text{asinh}\left(\frac{\sqrt{-\varDelta}}{2\bar{\rho}}\right)
\delta\rho \\
&+\frac{1}{2\pi}\int \,
\delta\theta\left(
\frac{\bar\rho^2}{K_\varDelta}\,
\text{asinh}\left(\frac{\sqrt{-\varDelta}}{2\bar\rho}\right)+
\frac{\varDelta}{8}
\right)
\delta\theta+\dots\;,\label{exact_result}
\end{aligned}
\end{equation}
where the dots indicate higher-order terms, the integrals
extend over the spacetime volume and
we introduced the (non-local) operator $K_\varDelta$ containing the Laplace operator,
\begin{equation}
K_\varDelta
=\left(1-\frac{4\bar\rho^2}{\varDelta}\right)^{1/2}\;.\label{nloperator}
\end{equation}
In a low-energy approximation we may perform 
the gradient expansion, which yields the simple expression
\begin{align} 
S_\mathrm{eff}&=
V U_\mathrm{eff}(\bar{\rho})\nonumber\\
&+\frac{1}{2\pi}\int
\left(\delta\rho^2+\frac{(\nabla\delta\rho)^2}{12\bar{\rho}^2}
-\frac{(\varDelta\delta\rho)^2}{120\bar\rho^4}+\dots\right)
\label{final_result}
\\
&+\frac{1}{16\pi}\int
\left((\nabla\delta\theta)^2-\frac{1}{3\bar{\rho}^2}
(\varDelta\delta\theta)^2+\dots\right)+\dots\nonumber\;,
\end{align}
containing the standard kinetic terms plus higher derivative terms.
The first term under the first integral is just the
second-order term in the expansion of $U_\mathrm{eff}(\bar\rho+\delta\rho)$
in powers of $\delta\rho$. Thus, up to second order
the effective action for  $\rho=\bar{\rho}+\delta\rho$ and $\delta\theta$
at low energies has the form
\begin{align} 
S_\mathrm{eff}&=
\frac{1}{4\pi}\int d^2x\,\rho^2\Big(\log\frac{\rho^2}{\bar\rho^2}-1\Big)\nonumber\\
&+\frac{1}{24\pi}\int d^2x\,
\Big(\frac{(\nabla\rho)^2}{\bar{\rho}^2}
-\frac{(\varDelta\rho)^2}{10\bar\rho^4}\Big)
\label{final_result3}
\\
&+\frac{1}{16\pi}\int d^2x\,
\Big((\nabla\delta\theta)^2-\frac{1}{3\bar{\rho}^2}
(\varDelta\delta\theta)^2\Big)+\dots\;,\nonumber
\end{align}
where we inserted the explicit form of the effective potential
at zero temperature and
the dots indicate cubic and higher-order terms and higher
derivative terms.
We see explicitly that $\rho$ describes a massive field
and $\delta\theta$ a massless would-be Nambu-Goldstone mode.
At large $\Nf$ the latter decouples from the system
while at finite $\Nf$ it destroys perfect long-range order.

To study long-range correlations we can safely neglect 
contributions from the massive field and
obtain, for large but finite $\Nf$, the valid approximation
\begin{equation}
C(x) 
\approx \bar\rho^2 e^{-2\ii kx}\big\langle e^{\ii\delta\theta(t,0)-\ii\delta\theta(t,x)}\big\rangle\;.
\label{eq:corrBKT1}
\end{equation}
It holds information about the dominant wave numbers of typical
configurations in an ensemble.
Due to the logarithmic divergence in the correlator of the 
massless scalar field one finds for $x\to\infty$
\begin{equation}
\big\langle e^{\ii\delta\theta(t,0)-\ii\delta\theta(t,x)}\big\rangle \to \begin{cases}
	x^{-\frac{1}{\Nf}}&T=0\;,\\
	e^{-x/\xi_\beta}&T>0\;,\label{asywitten}
	\end{cases}
\end{equation}
such that in a BKT phase with quasi-long-range order
the amplitude of the oscillating correlator
decays fairly slowly, following a power law.
At finite temperature the correlation length, given by
\begin{equation}
\xi_\beta=\frac{2\Nf}{\pi T}\alpha\;,\quad \alpha=1+
2\sum_{n\geq 1}(-1)^n(n\beta\bar{\rho})K_1(n\beta\bar{\rho})\;,
\label{eq:corrBKT1b}
\end{equation}
where $K_1$ denotes a modified Bessel function of the second kind, is finite. The coefficient $\alpha$ increases monotonically with
the inverse temperature $\beta$ from $\alpha=0$ to $\alpha=1$.
This means that the correlation length 
diverges in the large-$\Nf$ limit or for $T\to 0$.

	\section{Lattice Field Theory Approach}
	\label{sec:lattice}
	\subsection{Objectives and observables}
\label{sec:objectives}

The previous discussion makes clear that we should 
not expect to see
SSB  in the cGN model with U$_A$(1) symmetry.
Indeed, there are stronger arguments against perfect long-range 
order in this model than in the GN model with $\Z_2$ symmetry.
However, the difference between a spontaneously broken and a BKT 
phase at zero temperature most 
likely appears on exponentially large length scales that cannot be reached in our lattice simulations, see, for instance, \cite{BH94}. It could very well happen that on physically relevant length scales one can hardly distinguish between quasi-long-range and perfect long-range order. Furthermore, we shall see that even the distinction between a massive symmetric 
phase and a BKT phase at low temperatures is non-trivial if one allows for contributions 
of the first excited state.

Either way we will find
striking similarities between the cGN model with only two flavors
and the model with $\Nf\to\infty$, which, for $\mu\neq 0$, 
shows SSB of translation invariance. If similar observations apply to more realistic models
in higher dimensions then this could be relevant for
the physics of compact neutron stars, heavy-ion collisions or 
condensed matter in small systems.

\begin{figure*}[t]
	\includegraphics[scale=0.4]{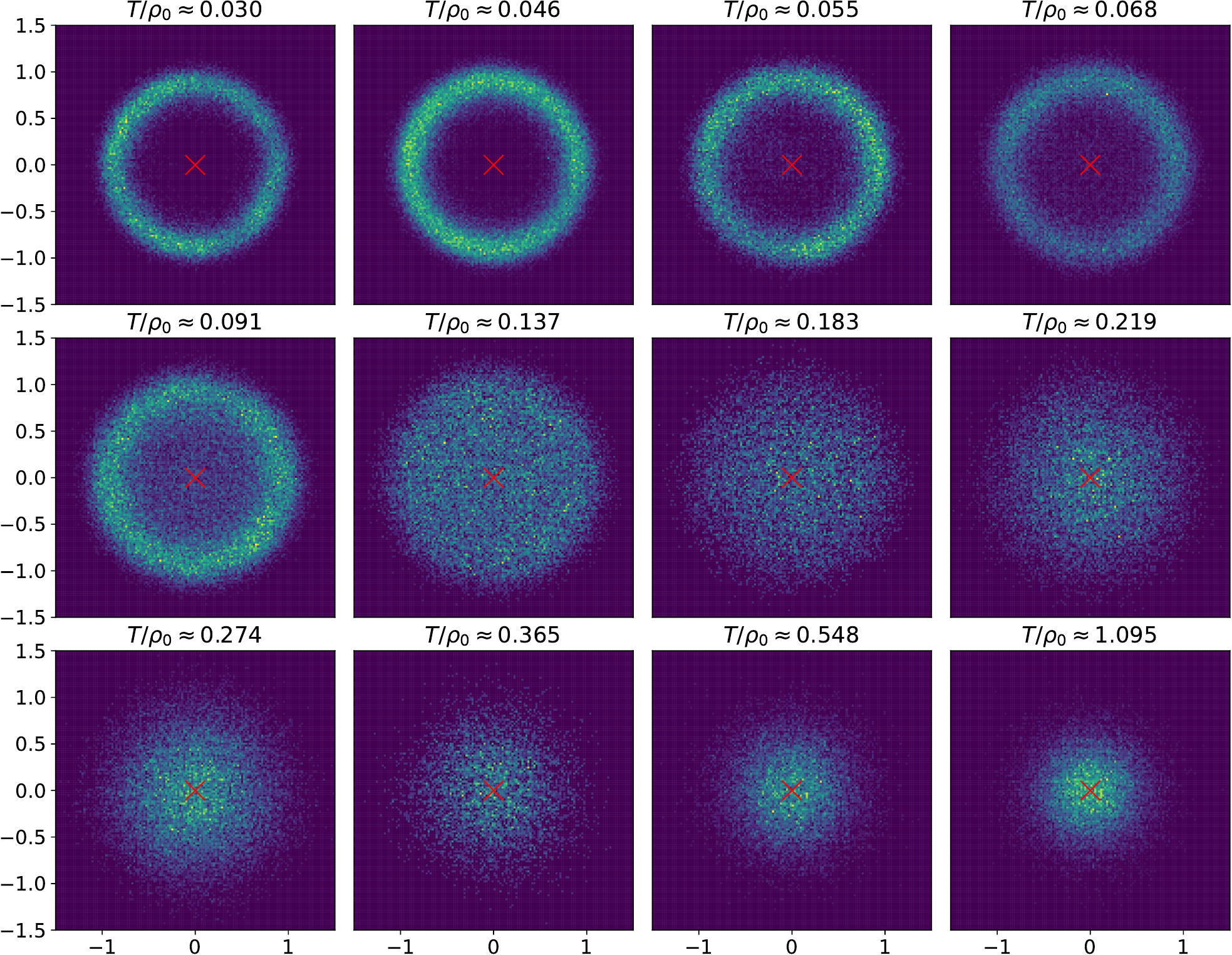}
	\caption{Distributions of $\frac{1}{\rho_0\Ns\Nt}\sum_{t,x}\Delta(t, x)$ in the complex 
		plane for 2 flavors, $N_s=63$, $a\approx0.46$ and $\mu=0$ at various temperatures. 
		The red crosses mark the origin and are included for visual clarity.}
	\label{fig:2dhistogram}
\end{figure*}

We shall see that for $8$ flavors the correlation function  
$C(x)$ in (\ref{eq:deltastar_delta_correlator})
has the form (\ref{eq:corrBKT1}) predicted by the effective low-energy Lagrangian
and can be hardly distinguished from  the large-$\Nf$ result (\ref{eq:Cspiral}).
For example, at low temperature its discrete Fourier transform $\mathcal{F}[C](k)$
is peaked at the dominant wave number
\begin{equation}\label{eq:kmax1}
\kmax =\frac{1}{2} \underset{k}{\argmax}\, |\mathcal{F}[C](k)|\;,
\end{equation}
which for large $\Nf$ is given by the chemical potential,
\begin{equation}\label{eq:kmax}
	\kmax\stackrel{\Nf\to\infty}{\longrightarrow}\mu\;,
\end{equation}
while for $\Nf=2$ we find
$\kmax<\mu$. Notice that we have included a factor of $1/2$ in \eqref{eq:kmax1}, in line with the introduction of $k$ in \eqref{eq:chiral_spiral} as half the 
wave number. We will use this convention for $k$ and $\kmax$ 
in the remainder of this work.

The spatial correlation function $C(x)$
also encodes the distinction between the
massive symmetric and BKT phases in 
its decay properties,
\begin{equation}
	|C(x)| \stackrel{x\to\infty}{\longrightarrow} 
	\begin{cases}
	e^{-x/\xi_\beta}&\text{massive symmetric}\;,\\
	x^{-\frac{1}{\Nf}}&\text{BKT}\;,\\
	\mathrm{const.}&\text{symmetry-broken}\;.
	\end{cases}\label{decaywitten}
\end{equation}
For a comparison we included the asymptotic behavior
in a symmetry-broken phase.
The temperature-dependent correlation length
$\xi_\beta$ was defined in \eqref{eq:corrBKT1b}.

\subsection{Lattice setup}\label{sec:setup}

We discretize two-dimensional Euclidean spacetime to a finite lattice with $N_s$ 
and $N_t$ lattice sites in the spatial and temporal directions respectively, 
such that $L=N_s a$ is the spatial extent, $T=1/N_t a$ is the temperature 
and $a$ denotes the lattice constant. 

In our simulations we employ chiral SLAC fermions \cite{DWY76_1, DWY76_2}, which discretize the dispersion relation in momentum space, leading to a non-local kinetic term in position space. They have proven advantageous over other fermion discretizations for low-dimensional fermionic theories, see e.g.~\cite{LPW20}. 
The use of SLAC fermions restricts $N_s$ to be odd and $N_t$ to be even. For further details we refer to sections 2.1 and 4.1 of
\cite{BKU08}. Note that our lattice setup is the same as
in \cite{LPW20}, with the only difference that
besides a scalar field $\sigma$ we now have an additional 
pseudo-scalar field $\pi$ and both fields enter the complex
condensate field $\Delta$ via \eqref{eq:delta}.

For an easy comparison with the analytic results
we express physical quantities in units of the
expectation value
$\left\langle \rho \right\rangle$ at $T=\mu=0$, denoted
by $\rho_0$. This is analogous to the scale $\sigma_{0}$ in our previous 
studies \cite{LPW20, LPW20_1}. 
One should stress
that this neither assumes any form of 
symmetry breaking nor is in conflict with any no-go theorem because 
a non-vanishing expectation value of $\rho$ does not break any symmetry. 
Fig.~\ref{fig:2dhistogram} shows histograms
of $\sum_{\vx}\Delta(\vx)$ in the complex plane 
for $\mu=0$ and $12$ different temperatures. For these histograms 
we used ensembles with $\mathcal{O}(10^4)$ configurations each. 
We clearly observe that the distribution of $\Delta$
is angle independent or U$_A$(1) invariant. At low temperature it is ring shaped with its maximum
at $\rho>0$, while at high temperature it turns into
a Gaussian-like distribution and the maximum moves to $\rho=0$.

In the actual simulations, however, the quantity $\rho_0$ 
is surprisingly hard to determine. \aref{app:scale} sheds some light 
on the details of this procedure. In summary, we used\footnote{Notice 
that this is not identical to taking the absolute
values and MC averages of the distributions shown in Fig.~\ref{fig:2dhistogram}. 
For a more detailed discussion about the order of taking absolute 
values and averages, see App.~\ref{app:scale}}
\begin{align}
	\left\langle \rho \right\rangle \approx \frac{1}{\Ns N_\mathrm{MC}}\sum_{\tau, x}\Big|\frac{1}{\Nt} \sum_{t}\Delta^{(\tau)}(t,x) \Big|\;,
\end{align}
with $\tau=1,\dots,N_\mathrm{MC}$ enumerating
the Monte Carlo (MC) configurations. This yields a good signal at low temperatures 
where $\langle\rho\rangle$ is required.

For most of our simulations, we used one of three different spatial 
extents $\Ns=63, 127, 255$ and lattice constants $a\rho_{0}\approx 0.46$, $0.19$, $0.08$ in order to study both the 
continuum limit and the infinite-volume limit. We vary the temperature 
by changing the number of lattice points in the temporal direction, $\Nt$, at 
fixed $a$ and we vary $a$ by changing the coupling $1/g^2$ in \eqref{eq:lagrangian}. 
For these lattices we map out phase diagrams in the $(T,\mu)$ 
plane. More details as well as a table summarizing all parameter settings 
are given in \aref{app:parameters}.

Experience with interacting fermion models teaches us \cite{LPW20} that
systems exhibiting (quasi-)long-range inhomogeneous structures can have 
rather long thermalization times when 
running simulations with randomly generated initial configurations, 
e.g., using a Gaussian distribution. As a way to counteract this problem, 
we employ a different approach for the majority of results presented in 
this work and perform a systematic "freezing out" in the following way: 
Starting at high temperatures with $\Nt\ll\Ns$, where thermalization 
times are not an issue, we generate at least $1000$ configurations to 
ensure proper thermalization. We then map the last of these configurations
to a lower-temperature lattice with $\Nt'>\Nt$ by simply repeating the data in 
the temporal direction and use it as a seed configuration on the larger
lattice. This reduces the thermalization period (where no measurements are performed)
if the temperature step is small. In our 
simulations we systematically approach lower and lower temperatures 
using this "freeze-out" procedure. This way we experienced significantly 
less "getting stuck" in some far from typical configurations, although 
it could still not be completely prevented from happening.

A cross-check with thermalized results using standard Gaussian-distributed 
initial configurations yields equivalent results, with the "freezing" 
method having noticeably better thermalization properties and thus 
overall smoother results. As an additional cross-check we also 
performed the inverse procedure, i.e., a "heating", for a handful 
of parameters in order to exclude any hysteresis effects caused by 
the "freezing" method. 

\begin{figure}[h]
	\includegraphics[width=\linewidth]{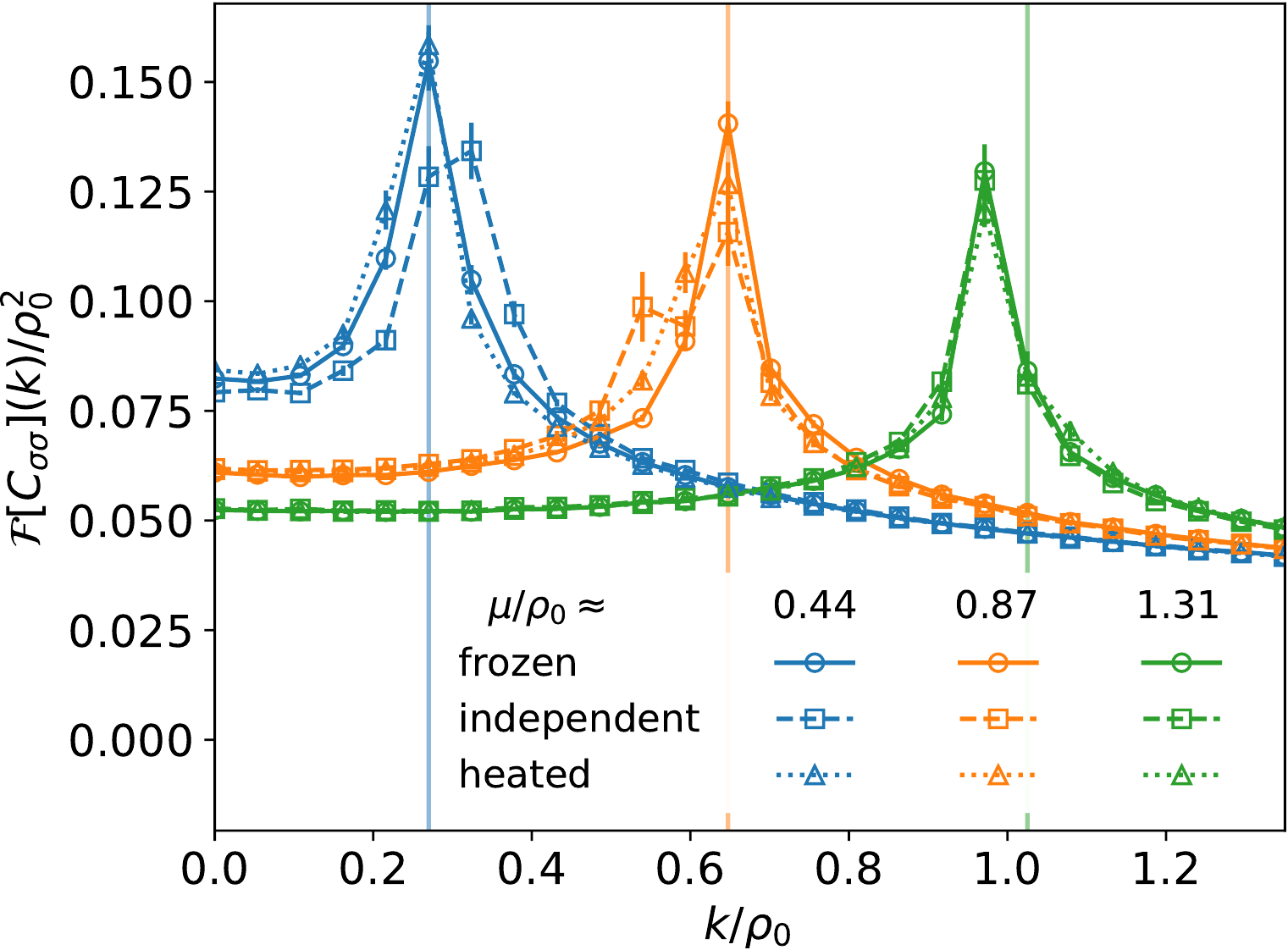}
	\caption{
	The Fourier transform of $C_{\sigma\sigma}(x)$ as a function of $k$ for the three different methods 
	mentioned in the main text on a $48\times127$ lattice for $a\rho_0\approx0.46$ and different $\mu$. 
	The temperature is the second lowest considered, i.e. we compare several 
	"freezing" steps with a single "heating" step. 
	The vertical lines indicate the maxima of the "frozen" results at the 
	lowest temperature.}
	\label{f:hysteresis}
\end{figure}

As can be seen from \fref{f:hysteresis}, where 
we show the Fourier transform of $C_{\sigma\sigma}(x)$ (to be defined in \eqref{eq:spatial_correlators}) computed via each of the
three methods, 
the "frozen" and "heated" results agree very well, indicating that 
hysteresis effects are negligible. The fact that the "independent" 
results, i.e. the ones obtained by using Gaussian-distributed initial 
configurations, show some deviation is likely to be attributed to 
their lower statistics and worse thermalization properties. 

The vertical lines in \fref{f:hysteresis} show the peak positions, 
which were estimated with the "freezing" method, for the lowest 
temperature considered. 
We see that for the highest density ($\mu/\rho_0\approx1.31$ in
the figure) the peak positions of the two lowest temperatures differ.
This dependence on temperature is not seen in the large-$\Nf$ limit and 
is caused by bosonic fluctuations.

For small $\mu$ on our smallest lattice, where homogeneous configurations dominate, we furthermore compared with a "cold start", which amounts to starting the simulation from $\Delta^{(0)}(t,x)=1+\ii$. Again we found matching results except for the lowest temperatures where the cold start is expected to suffer from severe autocorrelation problems. A detailed analysis of autocorrelation effects can be found in \aref{app:autocorrelation}.

\subsection{Lattice estimators}
\label{sec:lattice_estimators}

We have argued that spatial correlation functions are useful tools 
to probe for inhomogeneous phases since they avoid the destructive 
interference one would encounter when directly calculating chiral 
condensates on the lattice. We consider the two spatial correlators
\begin{eqnarray}\label{eq:spatial_correlators}
	\begin{aligned}
		C_{\sigma\sigma}(x)&=\frac{1}{N_tN_s}\sum_{t,y}\langle\sigma(t,y+x)\sigma(t,y)\rangle\;,\\
		C_{\sigma\pi}(x)&=\frac{1}{N_tN_s}\sum_{t,y}\langle\sigma(t,y+x)\pi(t,y)\rangle\;,
	\end{aligned}
\end{eqnarray}
where the sums extend over all lattice sites and $\langle\,\cdot\,\rangle$ 
denotes the Monte Carlo average. 
If these correlators show an oscillating behavior, 
one can infer the existence of inhomogeneities.
The unbroken U$_A$(1) symmetry (\ref{eq:chiral_symmetryb})
implies that for any temperature and chemical potential
\begin{equation}
C_{\sigma\sigma}(x)=C_{\pi\pi}(x)\quad \text{and}\quad
C_{\sigma\pi}(x)=-C_{\pi\sigma}(x)\;.\label{corr-rel}
\end{equation}
Also note that the fermion determinant is invariant when $\sigma$ and $\mu$
both change their signs, such that
\begin{equation}
\langle \sigma(\vx)\pi(\vy)\rangle_\mu=-\langle \sigma(\vx)\pi(\vy)\rangle_{-\mu}\;,
\label{eq:cormu}
\end{equation}
from which we conclude that
\begin{equation}
C_{\sigma\pi}(x)=C_{\pi\sigma}(x)=0\quad\text{for}\quad \mu=0\;.\label{eq:corzero}
\end{equation}
We see that additional correlators that arise from 
interchanging $\sigma\leftrightarrow\pi$ in \eqref{eq:spatial_correlators} 
are not independent and we refrain from using them in subsequent equations to save some space. 
In the measurements, however, we do not implement the 
symmetries (\ref{corr-rel}) and instead use all four correlators 
$C_{\sigma\sigma}$, $C_{\sigma\pi}$, $C_{\pi\sigma}$ and $C_{\pi\pi}$
in order to reduce statistical correlations. From 
\eqref{eq:deltastar_delta_correlator} one obtains
\begin{equation}
C(x)=2(C_{\sigma\sigma}(x)+\ii\,C_{\sigma\pi}(x))
\end{equation}
and the property (\ref{eq:corzero}) means that
$C$ is real for vanishing $\mu$.

In \cite{LPW20} we introduced 
the minimal value
\begin{equation}
	\label{eq:cmin}
	C_{\min} = \min_x C_{\sigma\sigma}(x)  \begin{cases}
		> 0& \text{homogeneous}
		\\
		\approx 0 &\text{symmetric}
		\\
		< 0& \text{inhomogeneous}
	\end{cases}
\end{equation}
to map out the entire phase diagram of 
the (discrete) GN model. This parameter is
negative if there is (quasi-) long-range order with 
oscillating $C_{\sigma\sigma}(x)$
and is also useful for discussing the physics of the 
chiral GN model. For the chiral model the choice of $C_{\sigma\sigma}$ 
might seem arbitrary but because of (\ref{corr-rel}) any quadratic
correlator of a linear combination of $\sigma$ and $\pi$ would serve the same purpose.

It is important to note that taking the minimum is a global operation that disqualifies this quantity as a \emph{local} observable.
Furthermore, this minimum might (and actually commonly will) be taken for small spatial separations $x$.
In such cases, $C_{\min}$ does not probe the long-range behavior of the system.

We estimate the dominant wave number $\kmax$ 
as given by \eqref{eq:kmax1}, but calculated from $C_{\sigma\sigma}$ instead of $C$.
Sometimes we quote results in terms of the integer-valued
dominant winding number
$\nmax$, related to $\kmax$ via
\begin{equation}\label{eq:nmax}
	\nmax = \frac{L}{\pi}\kmax\;.
\end{equation}

From analytical studies \cite{And05, Witten:1978qu} 
it is expected 
that the U$_A$(1)-invariant fermionic four-point function of
the GN model,
\begin{equation}
C_\mathrm{4F}(\vx;\vy)=\left\langle\bar{\psi}(1+\gamma_*)\psi(\vx)
\bar{\psi}(1-\gamma_*)\psi(\vy)
\right\rangle\;,\label{eq:wi1}
\end{equation}
at zero temperature and zero fermion density
should have a power-law behavior in the limit of 
large separations,
\begin{equation}
C_\mathrm{4F}(\vx;\vy)\sim
c\,\vert \vx-\vy\vert^{-\frac{1}{\Nf}}\;,\label{eq:wi3}
\end{equation}
where $c$ is some constant. 
Similarly to the spatial correlation functions (\ref{eq:spatial_correlators})
for the condensate fields we introduce the spatial 
correlation function for the $\Nf$ fermionic lattice fields,
\begin{equation}
C_\mathrm{4F}(x)=\frac{1}{N_tN_s}\sum_{t,y}C_\mathrm{4F}(t,y+x;t,y)\;.\label{eq:wi5}
\end{equation}
The asymptotic form (\ref{eq:wi3}) would imply a power-law decay
\begin{equation}
	 C_\mathrm{4F}(x)\sim c\,x^{-\frac{1}{\Nf}}\quad\text{for}\quad
	 x\gg 1\;.\label{eq:wi7}
\end{equation}
Dyson-Schwinger equations (see  \aref{app:4point})
relate
$C_\mathrm{4F}(x)$ to the spatial correlation functions of the
condensate fields,
\begin{equation}
	C_\mathrm{4F}(x) = -2\left(\frac{\Nf}{g^2}\right)^2\left(C_{\sigma\sigma}(x)
	+\ii\,C_{\sigma\pi}(x)\right)\;,\label{eq:wi9}
\end{equation} 
since the contact term in \eqref{apc:wi5} does not contribute
for large $x$. Since $C_{\sigma\sigma}$ and $C_{\sigma\pi}$ are
easily accessible in lattice simulations we
shall use this relation to study the infrared properties of
$C_\mathrm{4F}$. 
For $\mu=0$ the latter is real, see (\ref{eq:corzero}).

\begin{figure}[h]
	\includegraphics[width=0.95\linewidth]{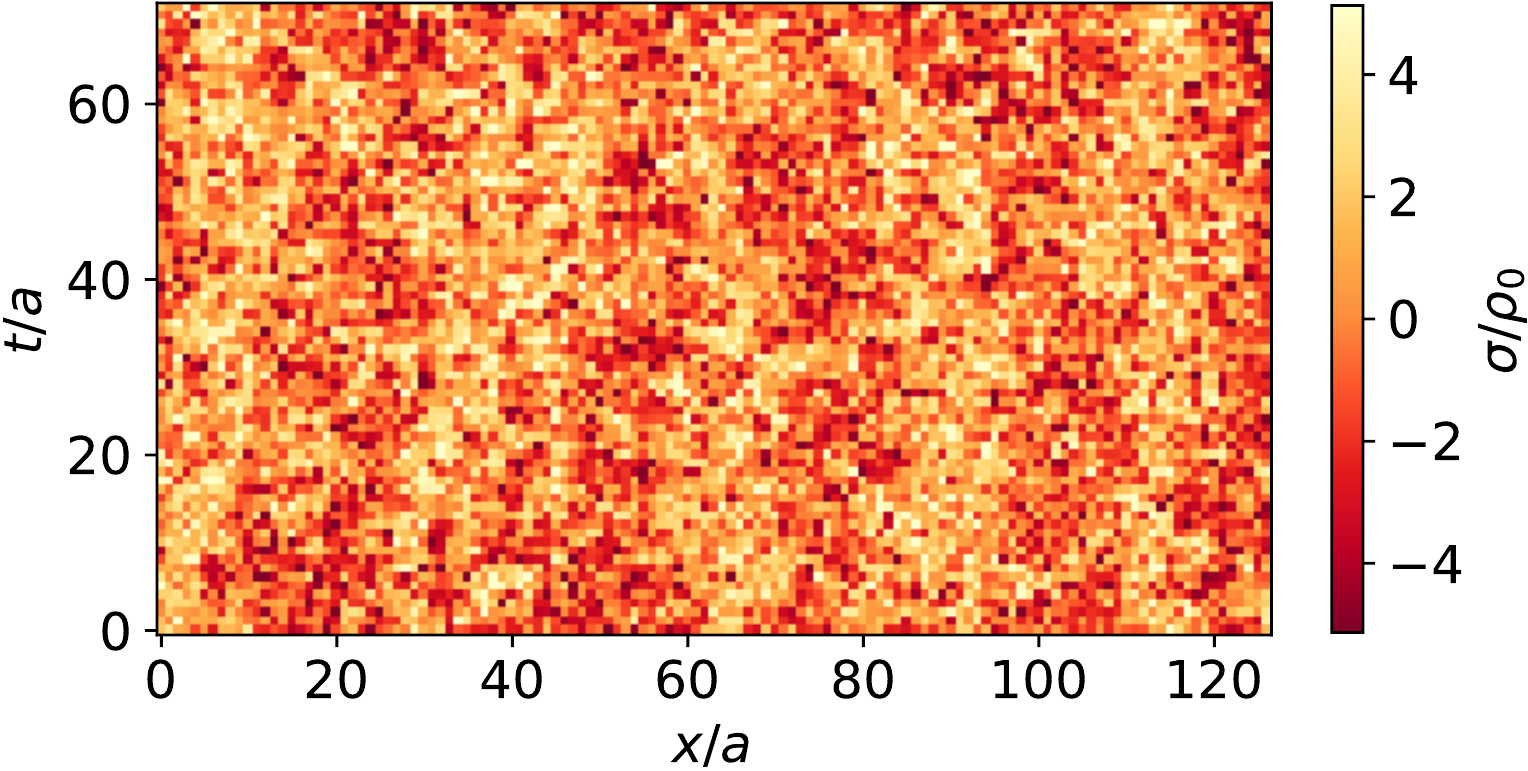}
	\caption{Typical configuration of $\sigma(\vx)$ on a $72\times127$ lattice 
	for $\mu/\rho_0\approx0.44$ and $a\rho_0\approx0.46$.}
	\label{f:sigma_configuration}
\end{figure}

From the effective low-energy approximation outlined 
in Sec. \ref{sec:perturbations} we expect that  the phase of 
the complex condensate field, $\theta=\arg(\Delta)$,
holds important information about the existence of inhomogeneous structures. 
We thus studied the space dependence of its expectation value, 
defined in the following way:
\begin{equation}
	\left\langle\theta(x)\right\rangle =
	\arg\left(\left\langle\bar{\Delta}(x)\right\rangle\right)\;,
	\label{eq:avtheta}
\end{equation}
where the bar indicates time averaging, i.e.,
\begin{equation}
	\bar{\Delta}(x)=\frac{1}{N_t}\sum_t\Delta(t,x)\,.
\end{equation}
We chose this (unusual) order of time- and MC averaging to 
suppress statistical uncertainties. Although the two averages
in (\ref{eq:avtheta}) do not commute the given prescription is
justified since the configurations are essentially constant in 
time direction, see \fref{f:sigma_configuration} for an example 
configuration of the $\sigma$ field.

	\section{Numerical Results}\label{sec:results}
	In previous studies of the discrete GN model \cite{LPW20,LPW20_1},
$2$, $8$ and $16$ flavors have been investigated with the focus on $\Nf=8$ 
in order to compare different types of chiral fermions\footnote{$\Nf=8$ is the
smallest number of flavors where naive fermions have no sign problem.}
and their suitability to investigate inhomogeneous phases. But
$\Nf=8$ is still close to $\Nf=\infty$ in the sense that on an 
intermediate scale quantum fluctuations away from the chiral spiral 
are suppressed.
To be more precise, if the BKT scenario were correct, 
then, for instance, in order to obtain a decay to half the 
amplitude a crude estimate using $C(x)\sim \vert x\vert^{-1/8}$ yields
\begin{equation}
\frac{C(x')}{C(x)}=\frac{1}{2}\quad\Rightarrow\quad \vert x'\vert = 256\,\vert x\vert
\end{equation} 
at the very least. Thus, in order to make any meaningful statements about such an amplitude decay 
we would require around $\mathcal{O}(10^3)$ lattice points 
at sufficiently small temperature (large temporal extent). This does 
not take into account severe autocorrelation problems,
finite-size effects and contributions from excited states that might all spoil the signal.
This crude estimate motivated us to study the long-range behavior for $\Nf=2$ 
in \cite{LPW20}, for which the same estimate yields feasible 40 lattice points.

\begin{figure*}
	\begin{subfigure}{.32\linewidth}
		\includegraphics[scale=0.37]{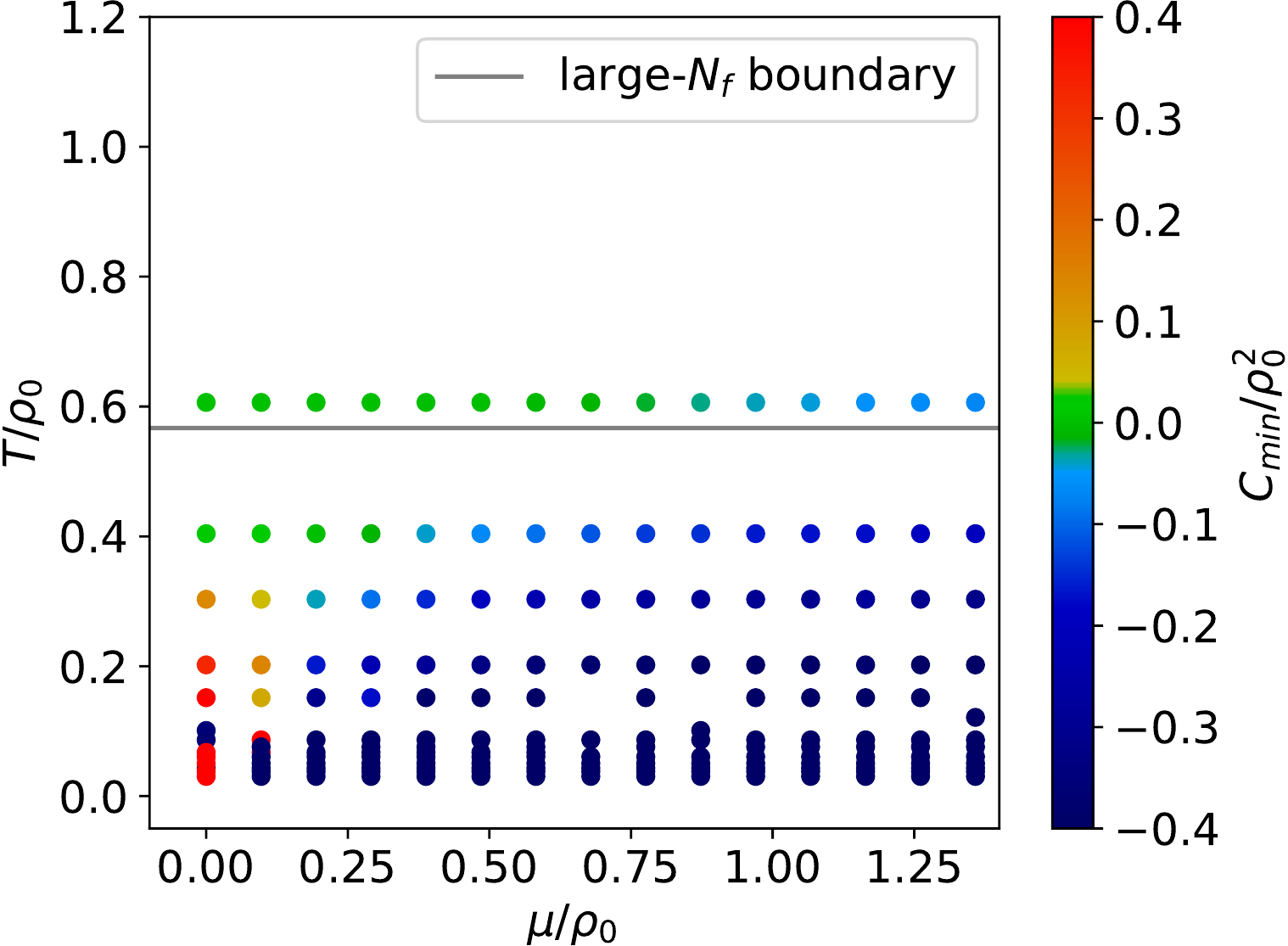}
		\caption{\label{f:nf8_pd}Phase diagram via $C_{\min}$ from \eqref{eq:cmin}.\\\phantom{ }}
	\end{subfigure}
	\hspace{0.1cm}
	\begin{subfigure}{.32\linewidth}
		\includegraphics[scale=0.45]{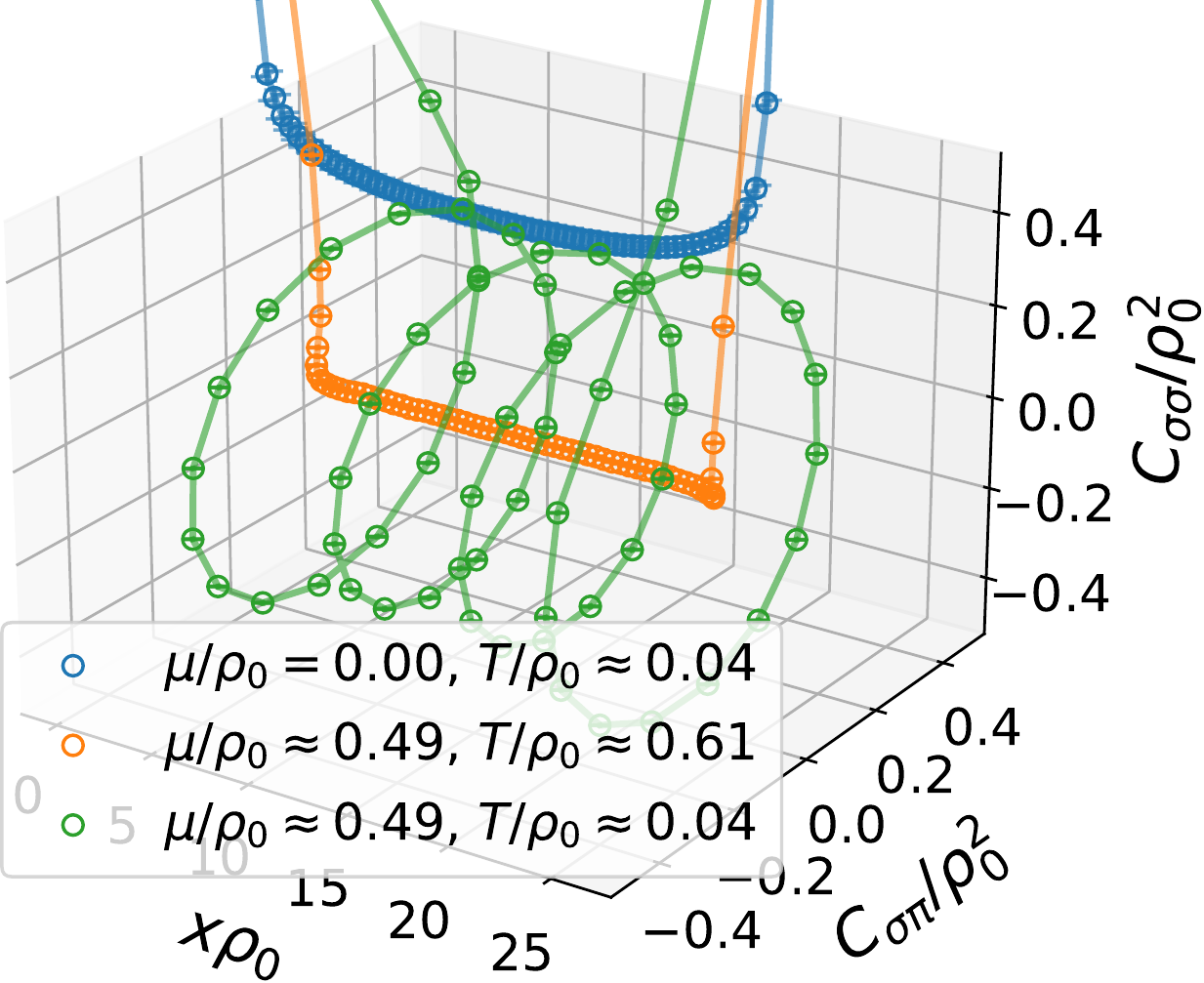}
		\caption{\label{f:nf8_correlators}Correlators $C_{\sigma\sigma}(x)$ and $C_{\sigma\pi}(x)$ from \eqref{eq:spatial_correlators} for various representative parameters.}
	\end{subfigure}
	\begin{subfigure}{.32\linewidth}
		\includegraphics[scale=0.37]{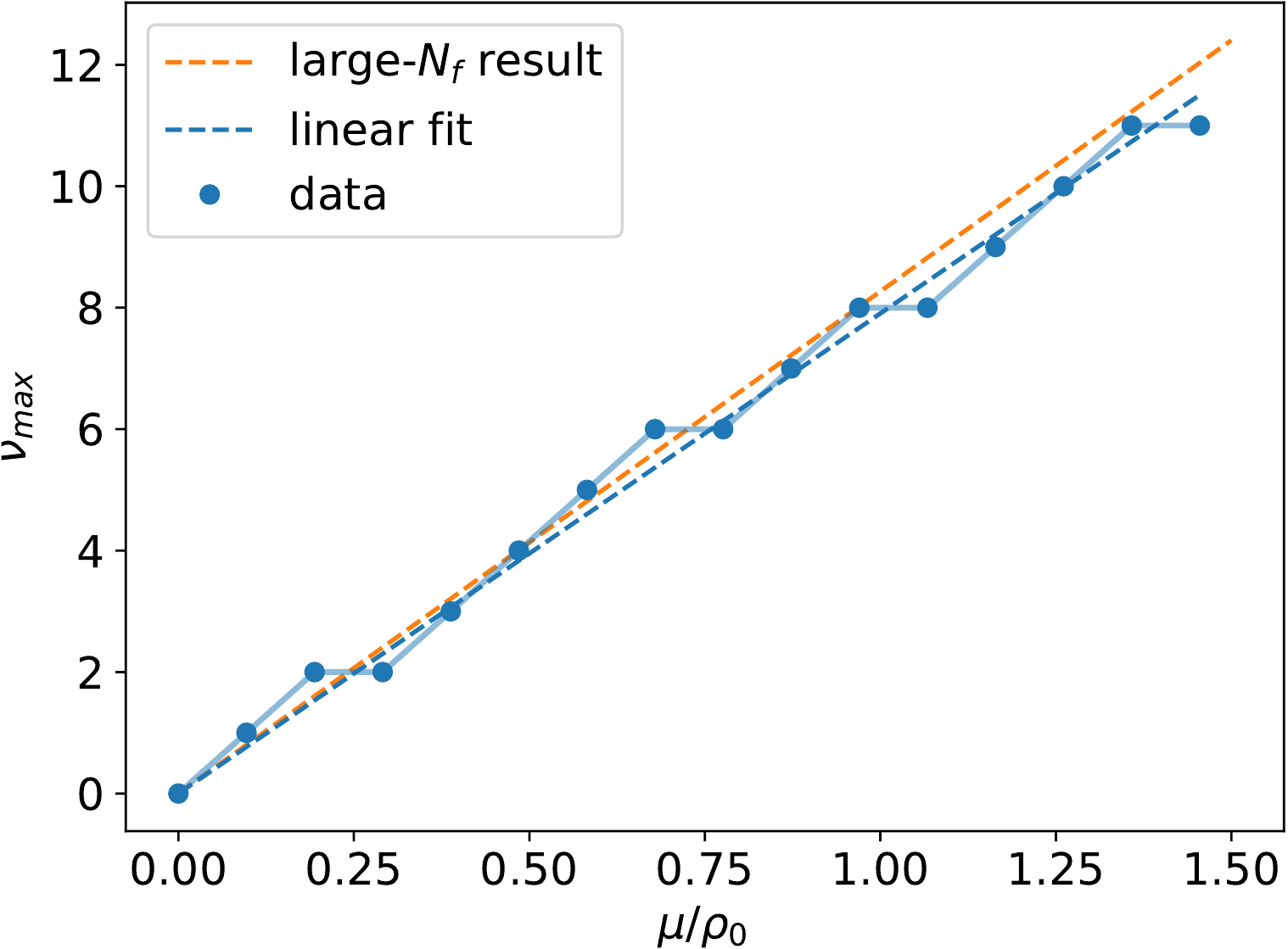}
		\caption{\label{f:nf8_nmax}Dominant winding number $\nmax$ from \eqref{eq:nmax} and \eqref{eq:kmax1} for $T/\rho_0\approx0.030$. The linear fit has a slope of $7.91\pm0.10$.}
	\end{subfigure}
	\caption{\label{f:nf8}Collection of results for $\Nf=8$, $\Ns=63$, $a\rho_{0}\approx0.41$.}
\end{figure*}

\subsection{Overview for \texorpdfstring{$\Nf=8$}{Nf=8}}
Although our focus is on $2$ flavors 
we performed one parameter scan in $(T,\mu)$ for $\Nf=8,\ \Ns=63$ and $a\rho_{0}\approx0.41$
in order to compare with results for the discrete GN model.
Some of our results are depicted in \fref{f:nf8}. 
\fref{f:nf8_pd} shows the phase diagram extracted from $C_{\min}$ (see \eqref{eq:cmin}), which 
is to be compared with \fref{f:large_n_phase_diagram} for infinite flavor number 
and is also the equivalent of Fig.~7 in \cite{LPW20}.
We see that the phase diagram agrees well with the large-$\Nf$ prediction for small chemical potential ($\mu< 0.5\rho_{0}$) and at least shows the predicted structure at larger $\mu$.

At vanishing
chemical potential $C_{\min}$ is positive for small temperatures, indicating 
predominantly homogeneous configurations with non-vanishing amplitudes. They relate
to the homogeneously symmetry-broken phase at large $\Nf$. In \fref{f:nf8_correlators} 
we see that in this regime $C_{\sigma\sigma}(x)$ is a positive and 
monotonically decaying function (blue curve) and $C_{\sigma\pi}(x)\approx 0$
in agreement with (\ref{eq:corzero}).
Raising the temperature we find a small temperature regime 
around $T\sim0.3\rho_0$ where we observe a sudden drop of the amplitude 
such that the $\mu=0$ data mimic a second-order phase transition. 
In the high-temperature regime the non-zero correlator $C_{\sigma\sigma}$ 
falls off even more rapidly.
This (would-be) transition temperature at $\mu=0$ is approximately equal to the one 
found in the discrete GN model in \cite{LPW20}. This was to be expected since in 
the large-$\Nf$ limit the GN and cGN models at vanishing 
chemical potential have the same critical temperature. It is also not surprising that 
for $\Nf=8$ the transition temperature is significantly lower 
than in the large-$\Nf$ limit (cf.\@ \eqref{eq:large_N_critical_temperature}), 
where quantum fluctuations are suppressed. 
The symmetric high-temperature regime at $\mu=0$ extends to non-vanishing 
chemical potential (orange curve in \fref{f:nf8_correlators}).

At low temperature and non-vanishing fermion density we can 
clearly confirm that the dominant contributions to the path integral come from 
chiral--spiral-like configurations. An example of this is shown in \fref{f:nf8_correlators} 
(green curve). Such configurations are the cause of the large region of negative values 
in \fref{f:nf8_pd}. The transition line to the region where oscillations are no longer 
dominant is roughly a line of constant temperature for small chemical potential ($\mu<0.5\rho_{0}$), as expected from the large-$\Nf$ solution.
For large chemical potential it tilts upwards unexpectedly, thereby enlarging the regime where inhomogeneities are found. 
This effect was also observed in \cite{LPW20} for $\Nf=2$ in the discrete GN model and is related to short- and intermediate-range phenomena that will be discussed later.
Nevertheless, the fact that we encounter it already for $\Nf=8$ strengthens the point that quantum fluctuations are much stronger in the chiral model compared to the discrete one.

For $\Nf=8$ the winding numbers (\ref{eq:nmax})
of the inhomogeneous configurations match 
the large-$\Nf$ expectation very well if one accounts for the discretization of the wave number due to the finite box size, as can be seen in \fref{f:nf8_nmax} (note that $\nmax$ is integer valued by definition).
As in \cite{LPW20_1} there is a tendency for the lattice data to lie slightly 
below the $\Nf=\infty$ expectation. The linear fit through the origin 
yields a slope of roughly $7.91$, which is lower than the large-$\Nf$ value $L/\pi\approx8.27$, but well within the expected accuracy of the large-$\Nf$ expansion $\mathcal{O}(1/\Nf)\sim 10\%$.

We remark that autocorrelations appear to be under control. However, due to limited statistics we cannot rule out the existence of another, larger, autocorrelation scale at low temperatures, see \aref{app:autocorrelation} for details.

\subsection{Deviations of \texorpdfstring{$\Nf\!=\!2$}{Nf=2} from the large-\texorpdfstring{$\Nf$}{Nf} limit}\label{sec:subsB}

\begin{figure}
	\includegraphics[width=\linewidth]{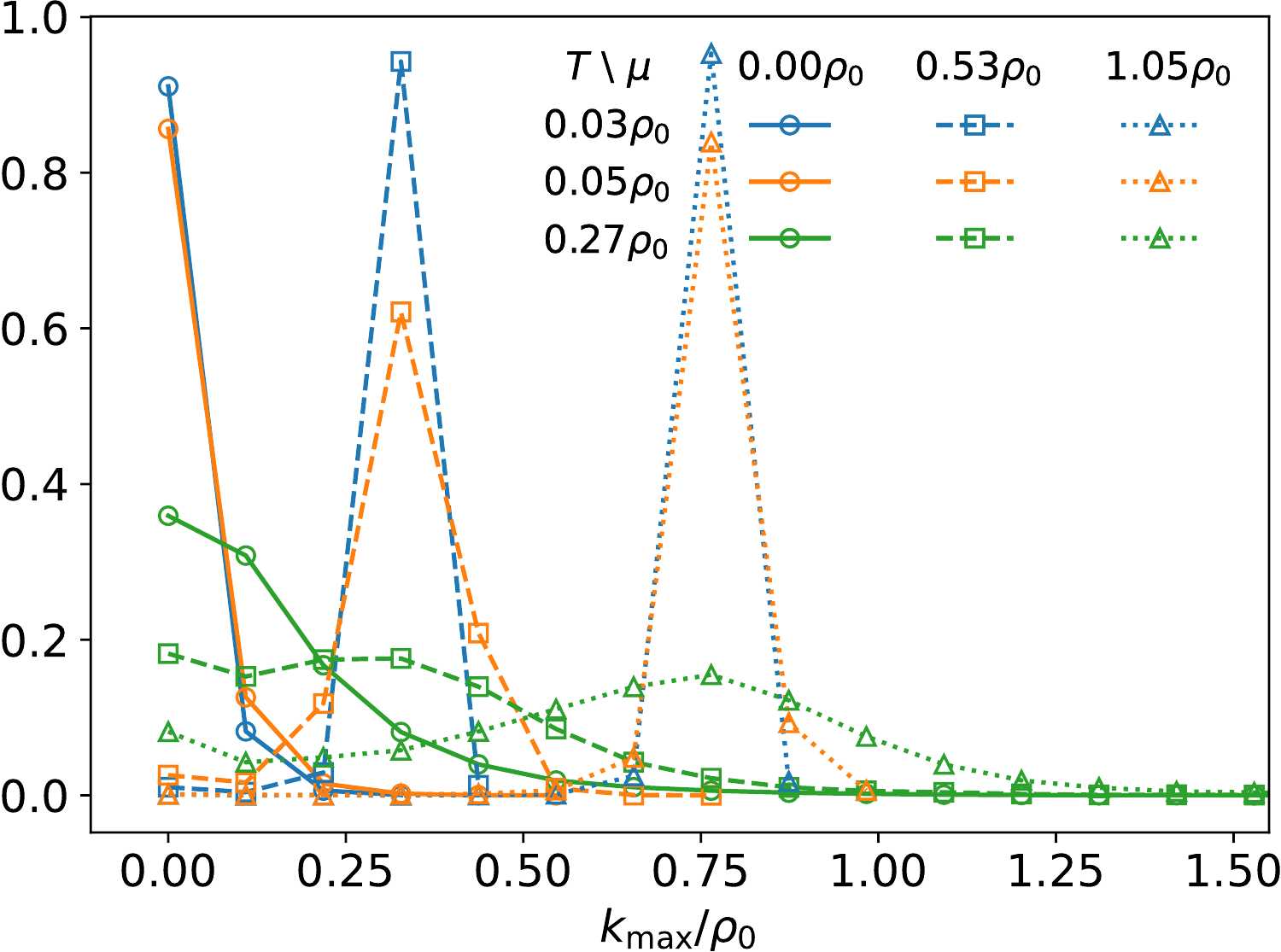}
	\caption{\label{f:nmax_hist}Histograms of $\kmax$ for $\Nf=2$, $\Ns=63$, 
	$a\rho_0\approx0.46$ and various values of temperature and 
	chemical potential.}
\end{figure}

\begin{figure*}
	\begin{subfigure}{.32\linewidth}
		\includegraphics[scale=0.47]{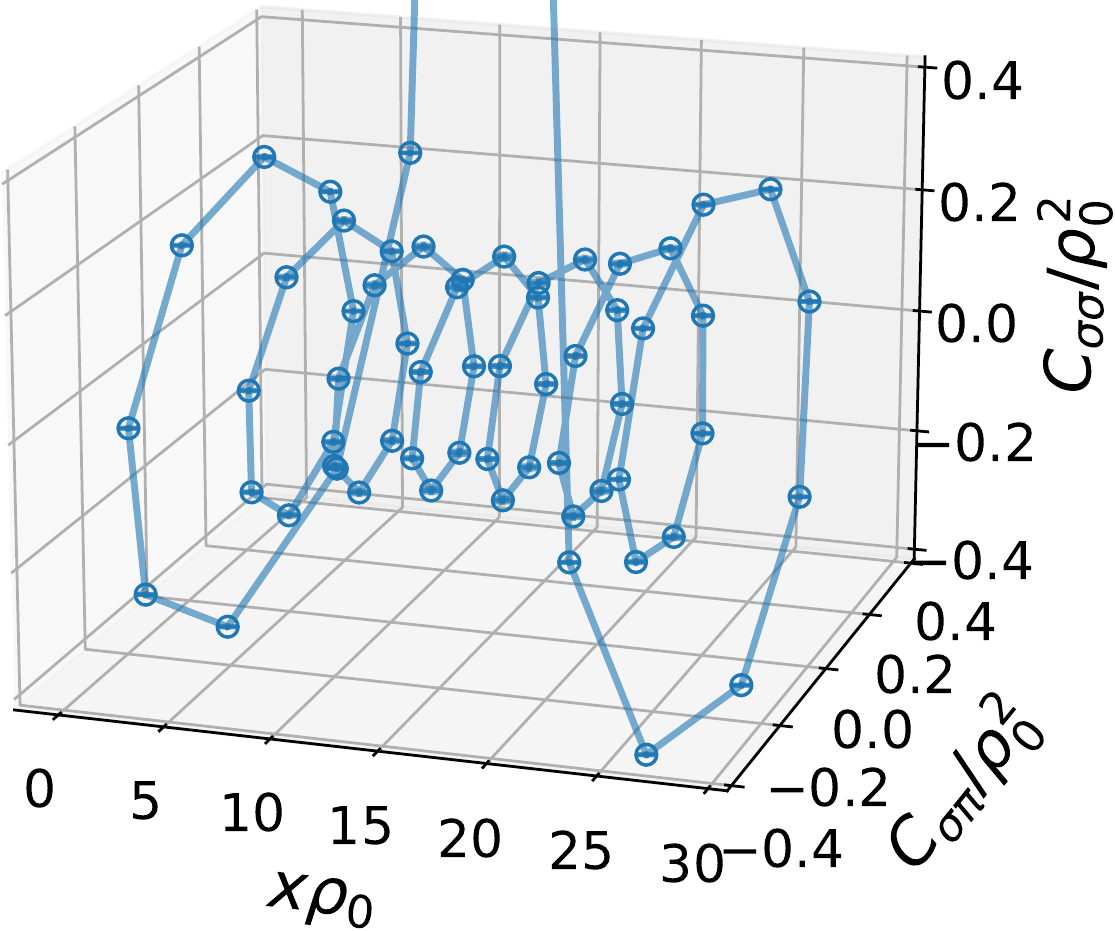}
		\subcaption{}
		\label{f:nf2_spiral}
	\end{subfigure}
	\begin{subfigure}{.32\linewidth}
		\includegraphics[scale=0.37]{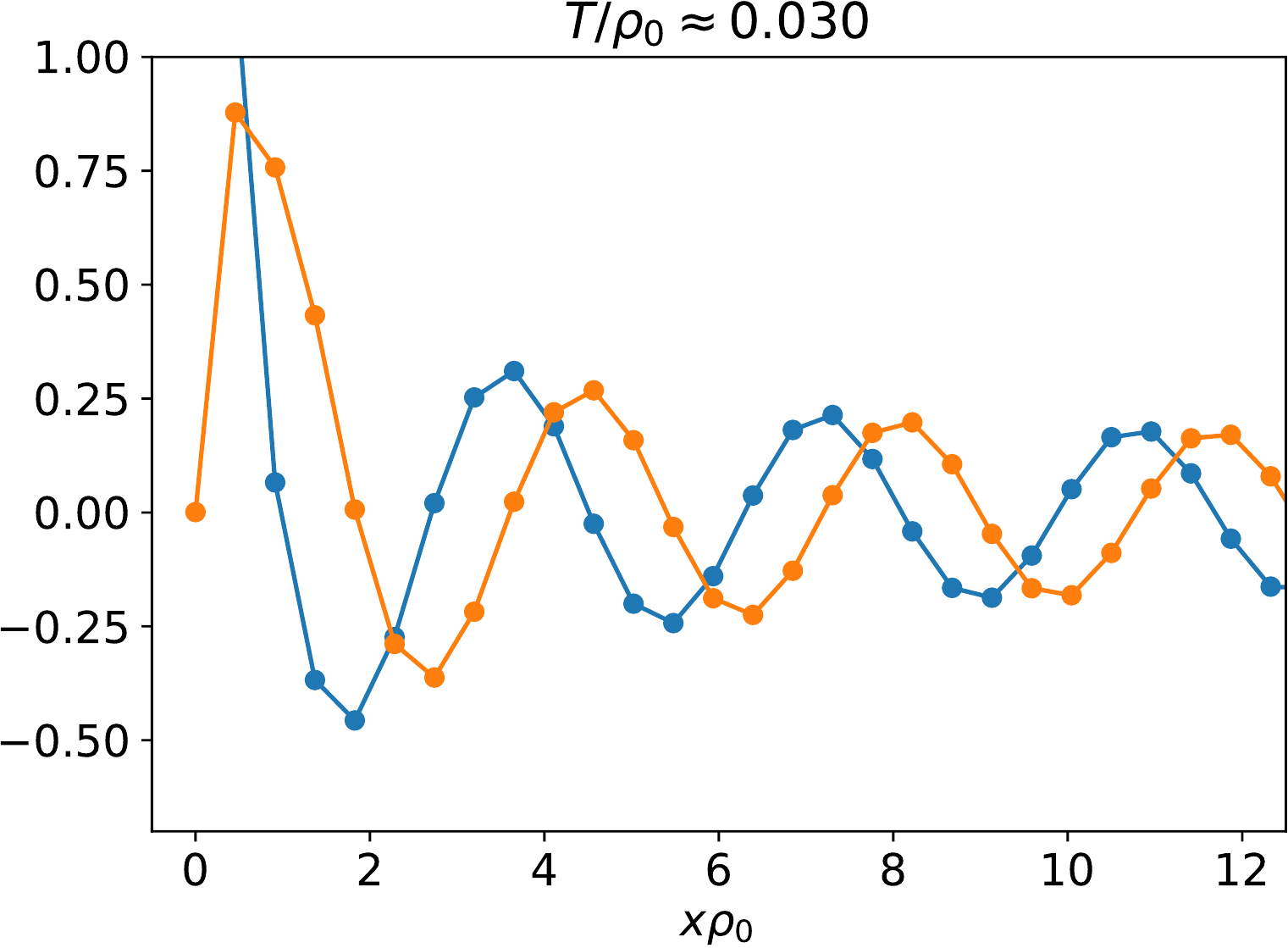}
		\subcaption{}
		\label{f:nf2_correlators_low_T}
	\end{subfigure}
	\begin{subfigure}{.32\linewidth}
		\includegraphics[scale=0.37]{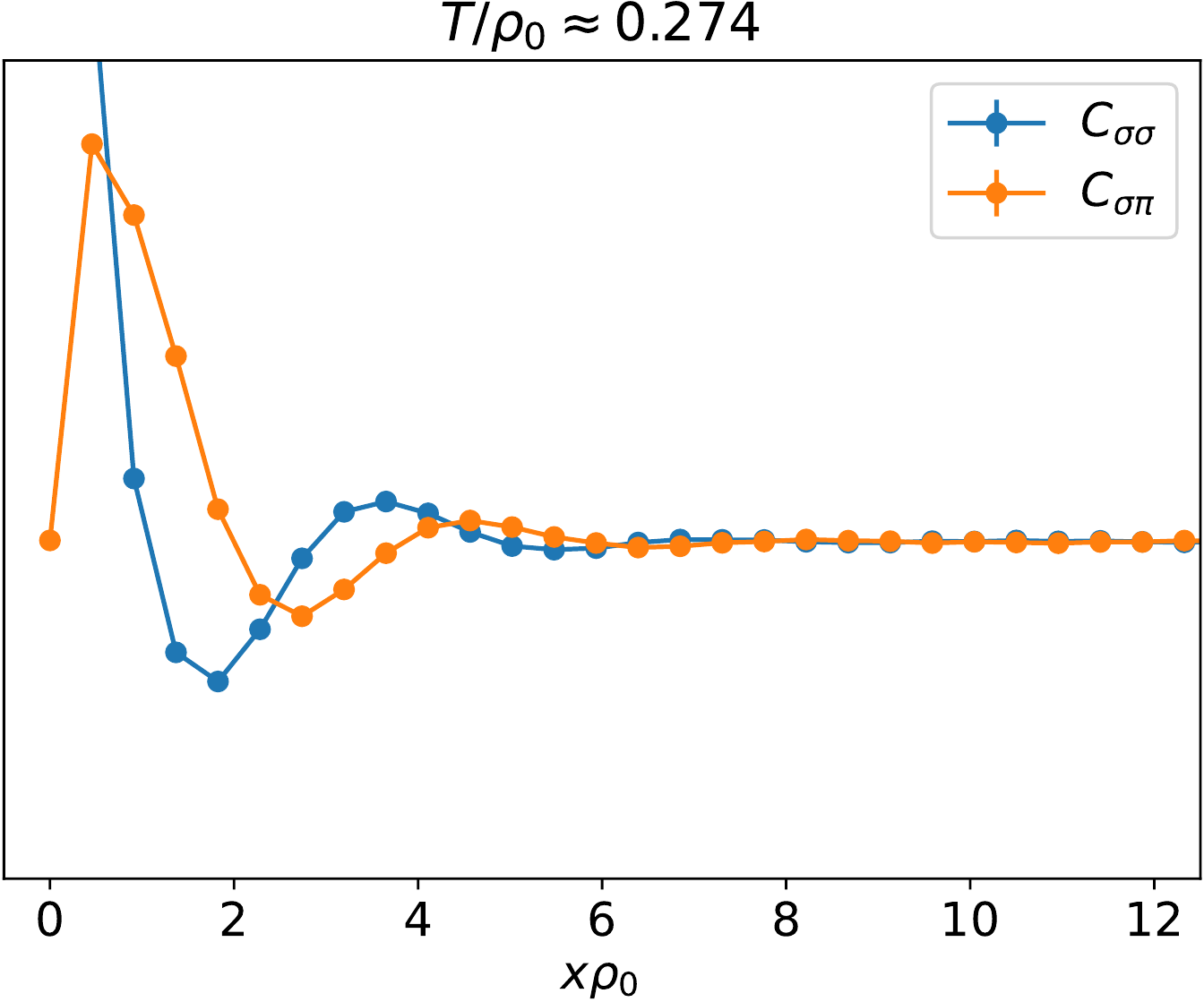}
		\subcaption{}
		\label{f:nf2_correlators_high_T}
	\end{subfigure}
	\caption{Correlators $C_{\sigma\sigma}$ and $C_{\sigma\pi}$  (from \eqref{eq:spatial_correlators})  for $\Nf=2$, $\Ns=63$, $\mu/\rho_0\approx1.14$ and $a\rho_0\approx0.46$. (a): 3D plot of the correlators at $T/\rho_0\approx0.030$ showing the chiral spiral. (b): The same data as in (a) but in 2D. (c) 2D plot of the correlators at higher temperature.}
	\label{f:nf2_correlators}
\end{figure*}

After discussing the results for $\Nf=8$, which in many ways confirm the large-$\Nf$ 
expectations, we now study the $2$-flavor cGN model for which we expect 
sizable deviations from the large-$\Nf$ solution.

To monitor the fluctuations in the system at finite temperature and
density, we measure the dominant wave number (\ref{eq:kmax1}) of
the equilibrium ensemble.
It characterizes important configurations for the given set
of control parameters and tells us which chiral spiral is favored
in the rough landscape defined by the effective action with 
its many local minima. This analysis presupposes that chiral spirals
are the dominant configurations even for $\Nf=2$ or that the non-dominant winding 
numbers are suppressed. We shall see that this is a valid assumption 
at small temperatures.

\fref{f:nmax_hist} shows such histograms for $3$ values of $T$ and
$3$ values of $\mu$.
As expected, the data show three distinct peaks, one for 
each value of $\mu$. At the lowest temperature and $\mu\neq 0$
the peaks are pronounced with over $80\%$ of the configurations sharing
the same dominant wave number. Increasing the temperature then broadens the peaks.  Concerning the question of spontaneous 
symmetry breaking, one should stress three features:
\begin{enumerate}
	\item\label{i:ft1} While the peaks flatten significantly for rising 
	temperature, 
	they do not vanish completely. At temperatures as high as $\sim 0.5\rho_{0}$ 
	we could still make out small bumps (not shown in \fref{f:nmax_hist}). Thus, even at these high temperatures the 
	system knows about the inhomogeneity arising from its finite density.
	\item\label{i:ft2} There is 
	no qualitative (or even sudden) change in these distributions that could characterize 
	a second-order phase transition. Instead the flattening of the peaks is a rather 
	smooth process.
	\item Even at low temperatures (e.g.~$T\approx 0.05\rho_{0}$), 
	well inside the would-be symmetry-broken regime, the contributions from concurrent 
	frequencies are significant (around $10-20\%$ in the example). The interference 
	of these contributions is the mechanism which, crucially, prevents a breaking of symmetry.
\end{enumerate}

\begin{figure}
	\includegraphics[scale=0.5]{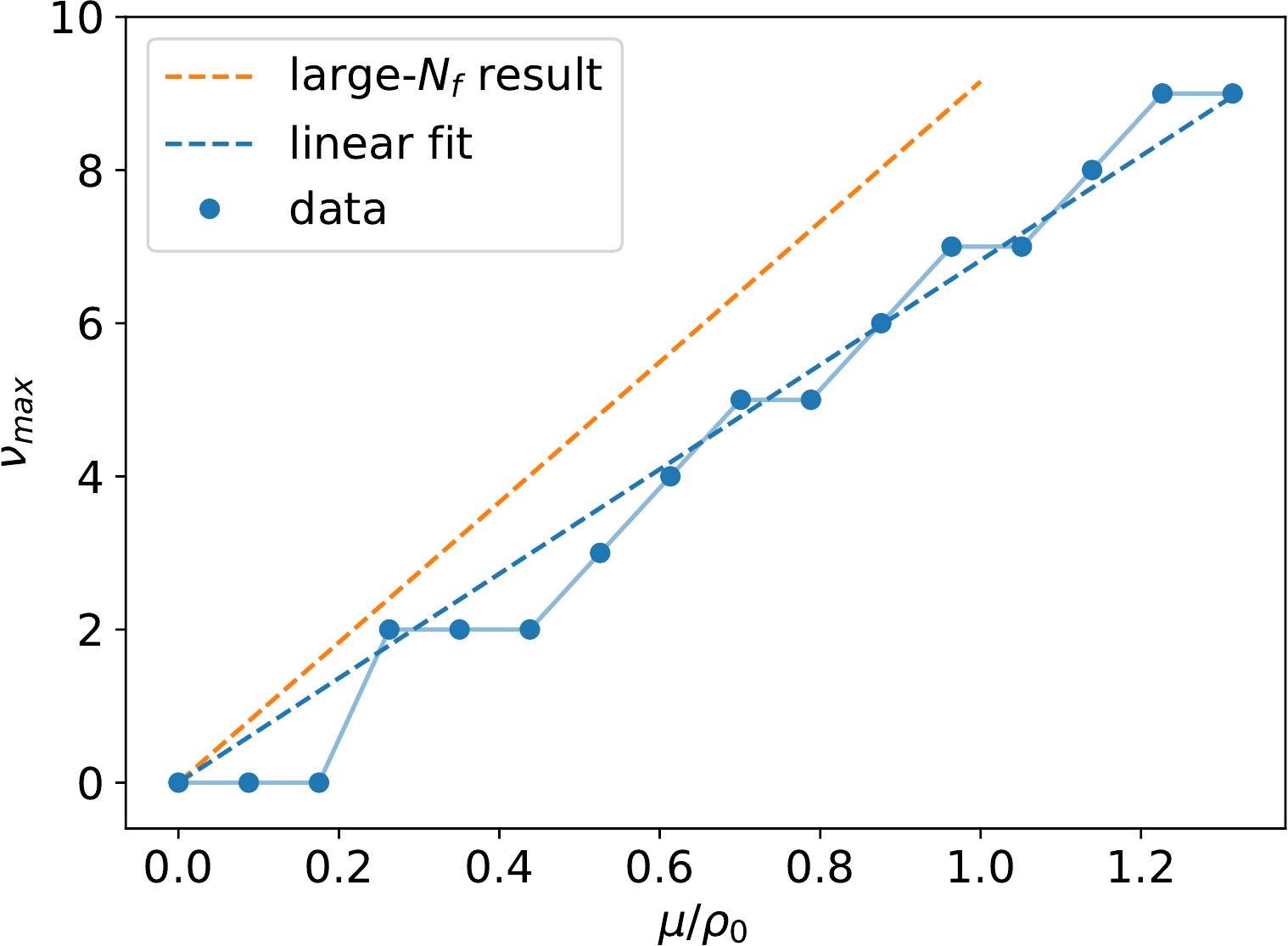}
	\caption{$\nmax$ as a function of $\mu$ for $\Nf$=2, $\Ns=63$, $a\rho_0\approx0.46$ and $T/\rho_0\approx0.030$. The linear fit has a slope of $6.82\pm0.17$.}
	\label{f:nf2_nmax}
\end{figure}

\begin{figure*}
	\begin{subfigure}{.333\linewidth}
		\includegraphics[scale=0.41]{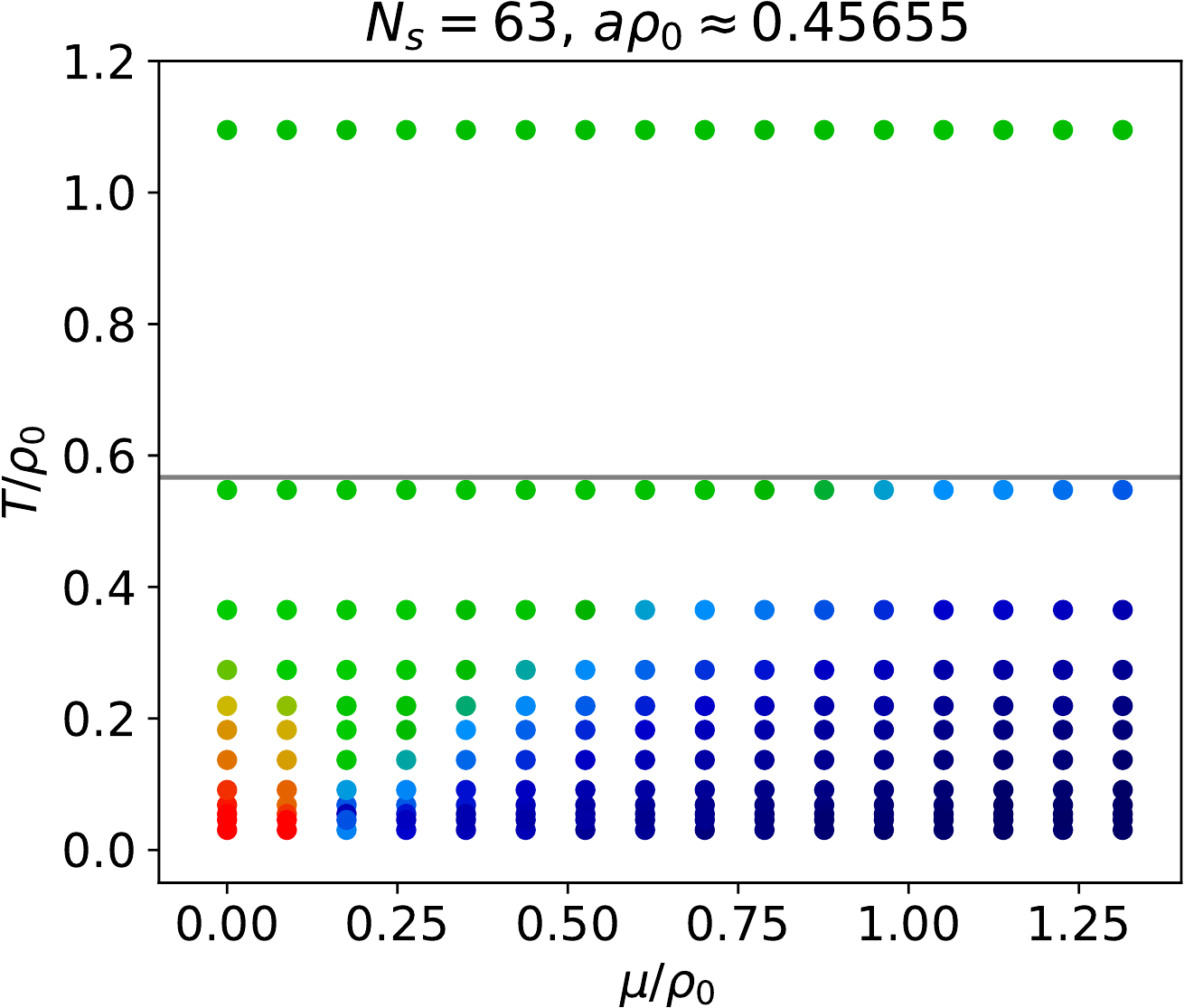}
	\end{subfigure}
	\hspace{-0.6cm}
	\begin{subfigure}{.333\linewidth}
		\includegraphics[scale=0.41]{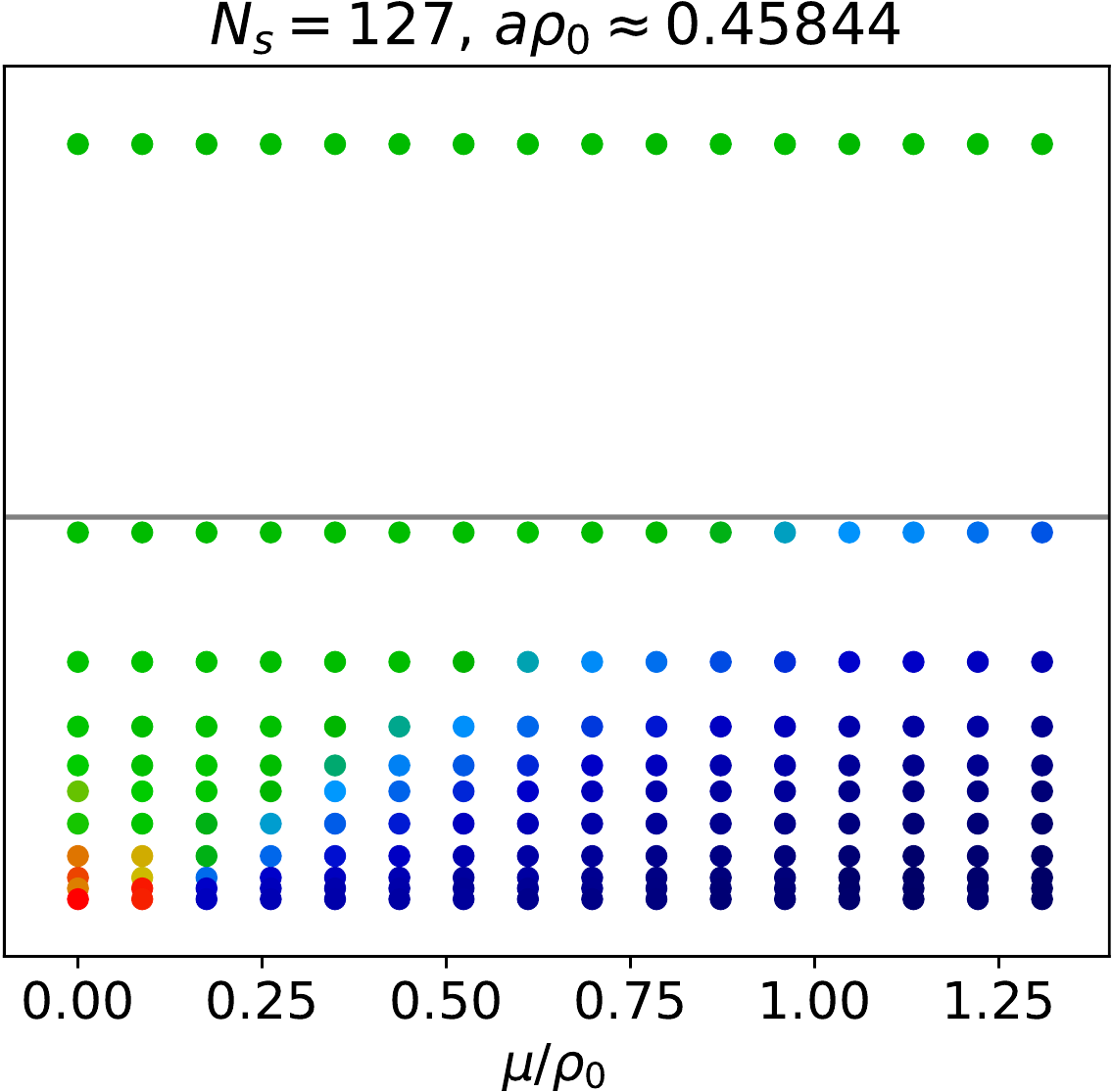}
	\end{subfigure}
	\begin{subfigure}{.333\linewidth}
	\hspace{-0.35cm}
		\includegraphics[scale=0.41]{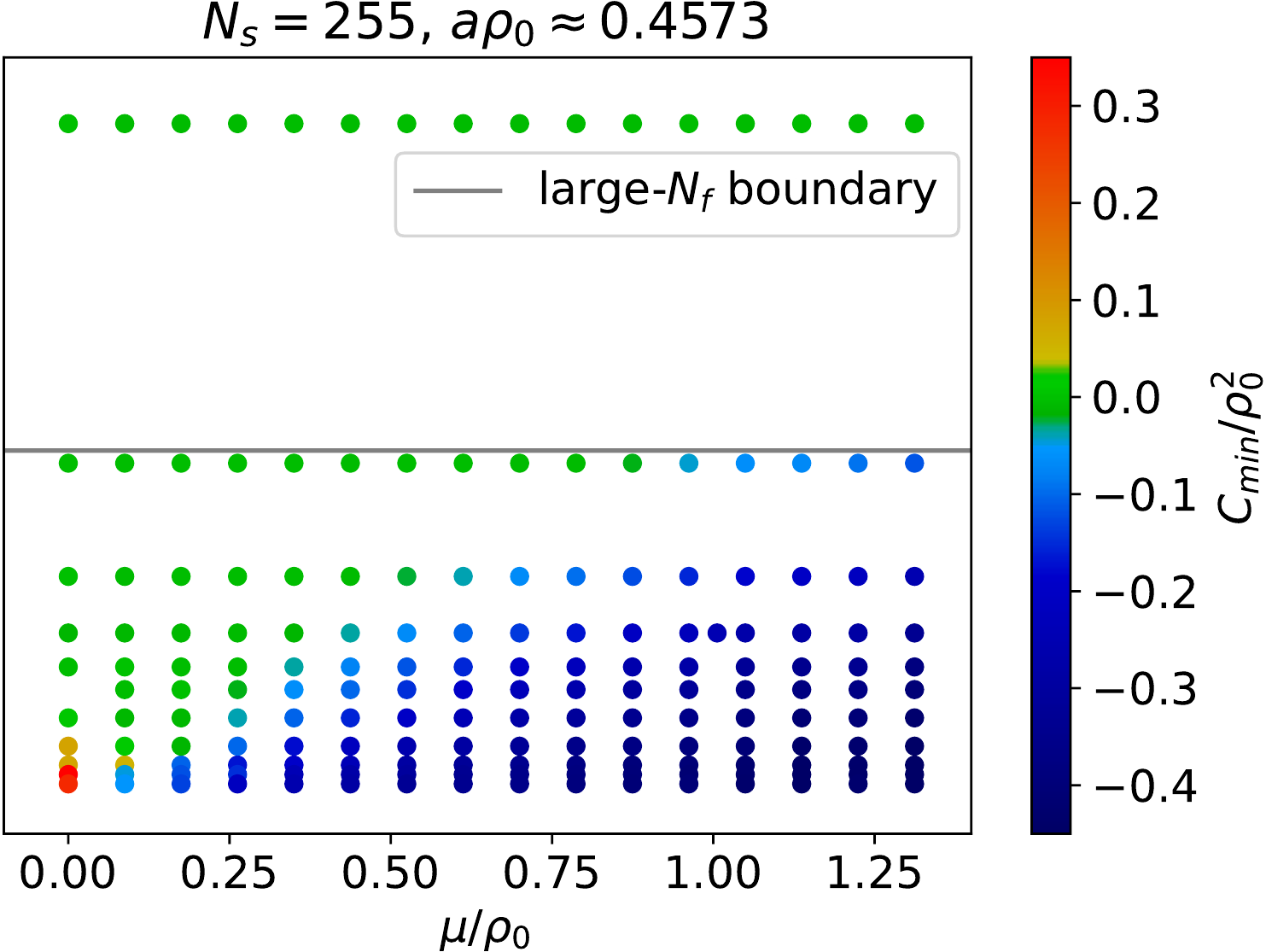}
	\end{subfigure}
	\caption{Infinite-volume extrapolation: phase diagrams for fixed lattice spacing and
	$N_s=63,127$ and $255$.}
	\label{f:pd_nf2_infinite_volume}
\end{figure*}

\begin{figure*}
	\begin{subfigure}{.333\linewidth}
		\includegraphics[scale=0.4]{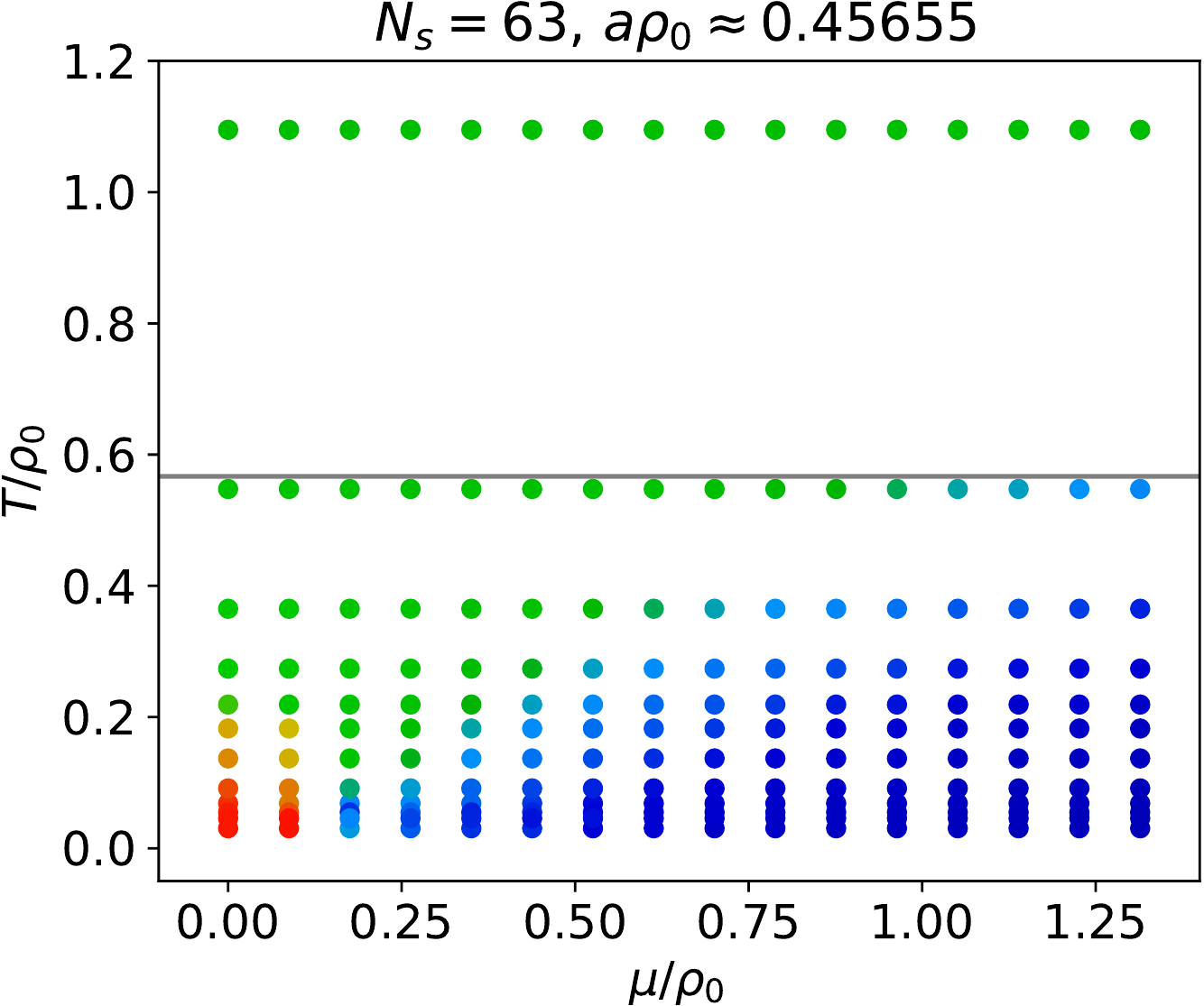}
	\end{subfigure}
	\hspace{-0.6cm}
	\begin{subfigure}{.333\linewidth}
		\includegraphics[scale=0.4]{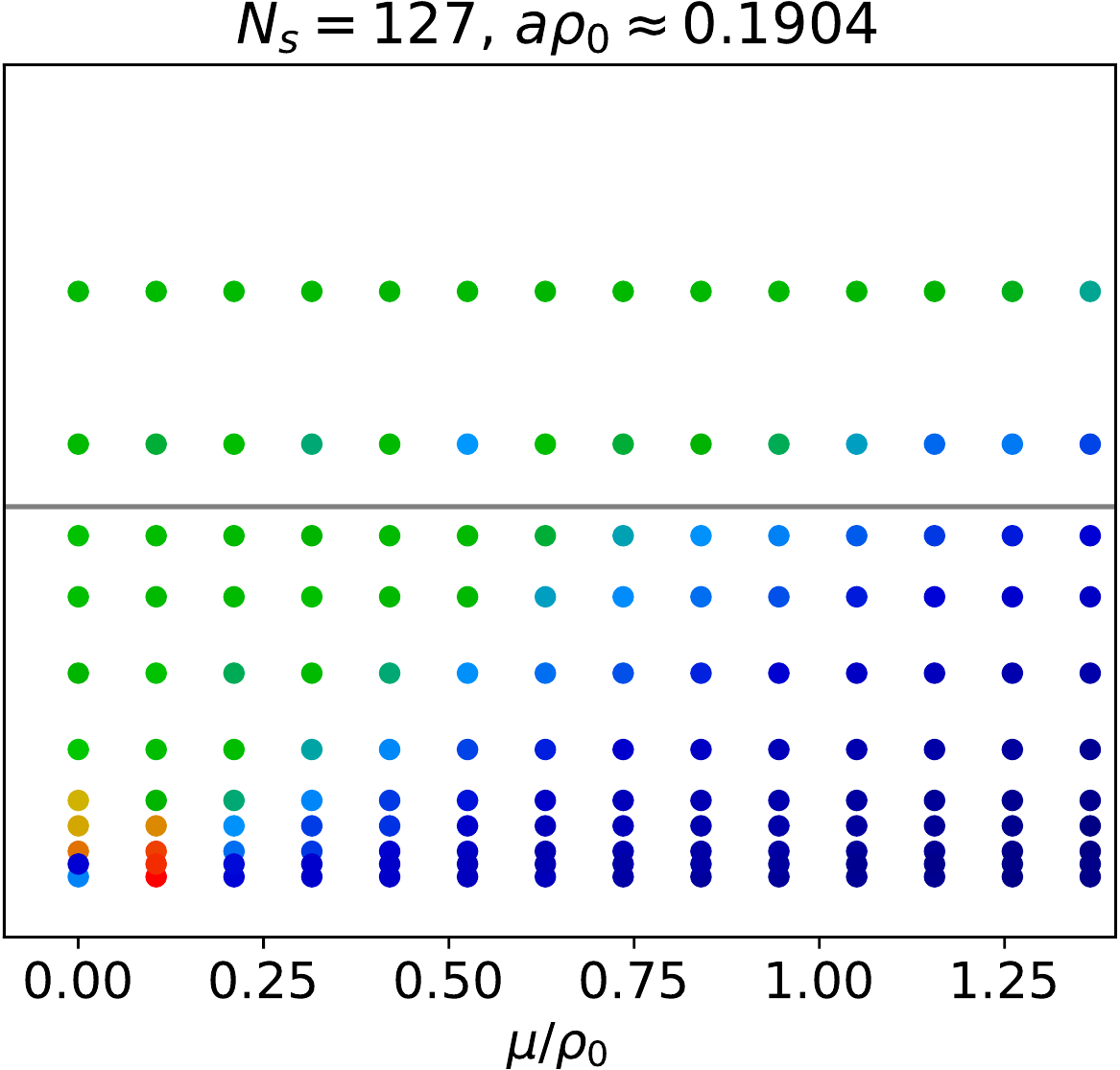}
	\end{subfigure}
	\begin{subfigure}{.333\linewidth}
	\hspace{-0.4cm}
		\includegraphics[scale=0.4]{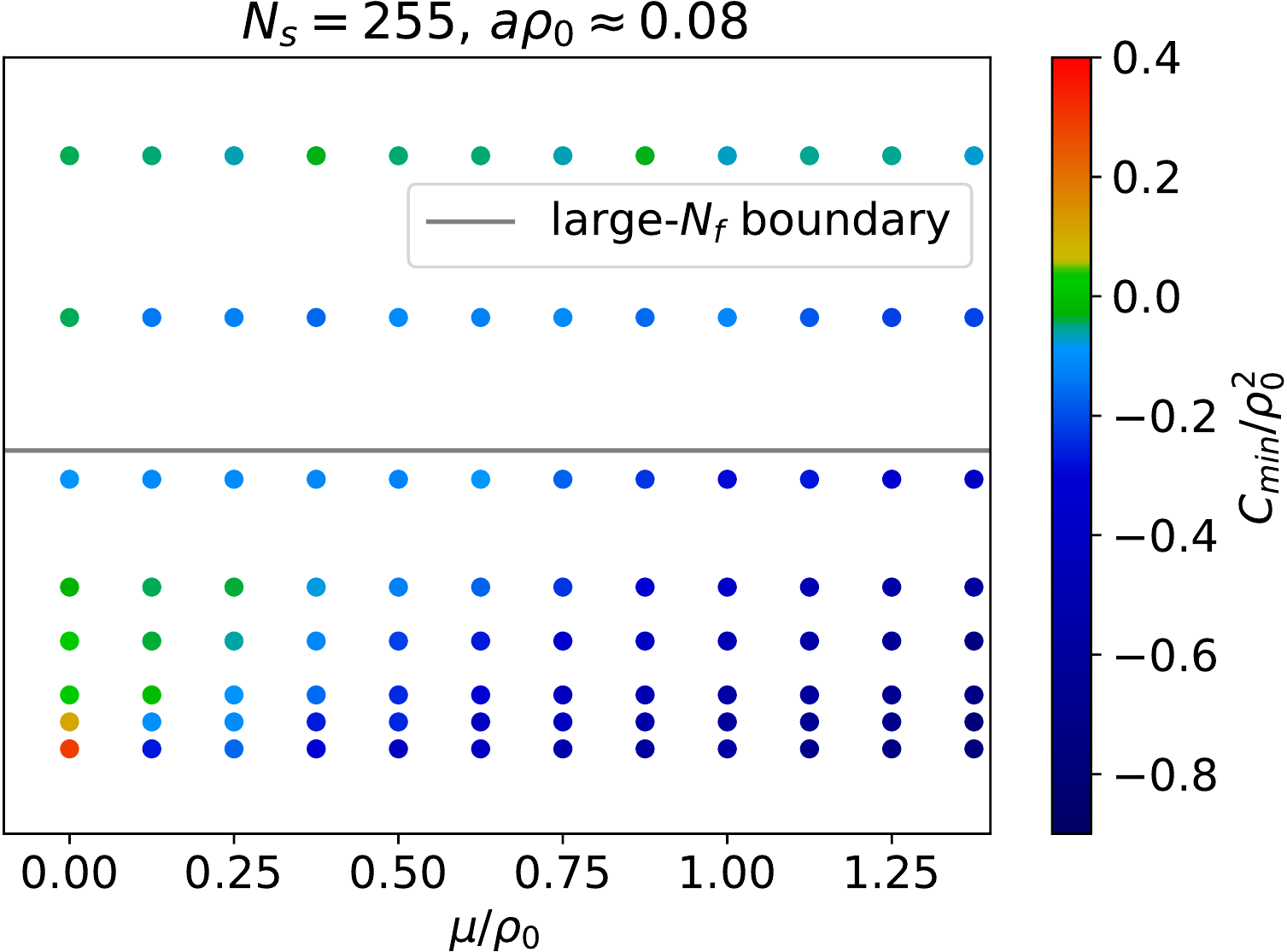}
	\end{subfigure}
	\caption{Continuum extrapolation: phase diagrams for decreasing
		lattice constant. The color coding is different from \fref{f:pd_nf2_infinite_volume}.}
	\label{f:pd_nf2_continuum}
\end{figure*}

The features \ref{i:ft1} and \ref{i:ft2} discussed above are 
clearly visible 
in the spatial correlators depicted in \fref{f:nf2_correlators}.
At low $T$ and non-vanishing $\mu$ we clearly observe remnants of 
a chiral spiral, see \fref{f:nf2_spiral}. From \fref{f:nf2_correlators_low_T}
we see that both correlators are oscillating with a phase shift of $\pi/2$.
This is the parameter region where there 
are sharp peaks in \fref{f:nmax_hist}. At higher temperature the peaks flatten and
the correlators show damped oscillations as shown in
\fref{f:nf2_correlators_high_T}. Notice, however, that even in this regime we still 
find $C_{\min}<0$, i.e., this is classified as a region of spatial inhomogeneities 
according to our definition. 
Here we observe a clear deviation from the large-$\Nf$ solution.
Since the oscillations in \fref{f:nf2_correlators_high_T} are only
seen on short scales we must be cautious when interpreting
a negative $C_{\min}$ as a signal for inhomogeneities. As already stressed before, a negative $C_{\min}$ ensures that there are oscillations on some length scale but this scale can be -- and certainly \emph{is} for large parts of the blue region of the phase diagram -- a short or intermediate one. Finally, at even higher temperatures one again finds 
correlation functions with $C_{\min}\approx0$, indicating a symmetric region. 

Similarly to $\Nf=8$ we determined the dependence of the dominant 
winding number in \eqref{eq:nmax} on the chemical potential 
and we present the results in \fref{f:nf2_nmax}. 
As expected, the (dominant) winding numbers for $\Nf=2$ deviate considerably 
from those in the large-$\Nf$ limit and those for $\Nf=8$, cf. \fref{f:nf8_nmax}.
One might conjecture that the winding numbers decrease with decreasing $\Nf$.

\subsection{Phase diagram for \texorpdfstring{$\Nf\!=\!2$}{Nf=2}}

One could expect a qualitatively different "phase diagram"
for the cGN model with $2$ flavors as compared to the large-$\Nf$ 
diagram depicted in \fref{f:large_n_phase_diagram}. In order to test this 
expectation we calculated $C_{\min}$ defined in \eqref{eq:cmin} on a
grid in the space of control parameters $\mu$ and $T$ on
lattices with $N_s=63$, $127$ and $255$ lattice points in the spatial 
direction. We studied both an infinite-volume extrapolation at 
(approximately) fixed
lattice constant $a\rho_0\approx 0.46$ and a continuum extrapolation 
at (approximately) fixed physical volume. 

\begin{figure}[h]
	\includegraphics[width=\linewidth]{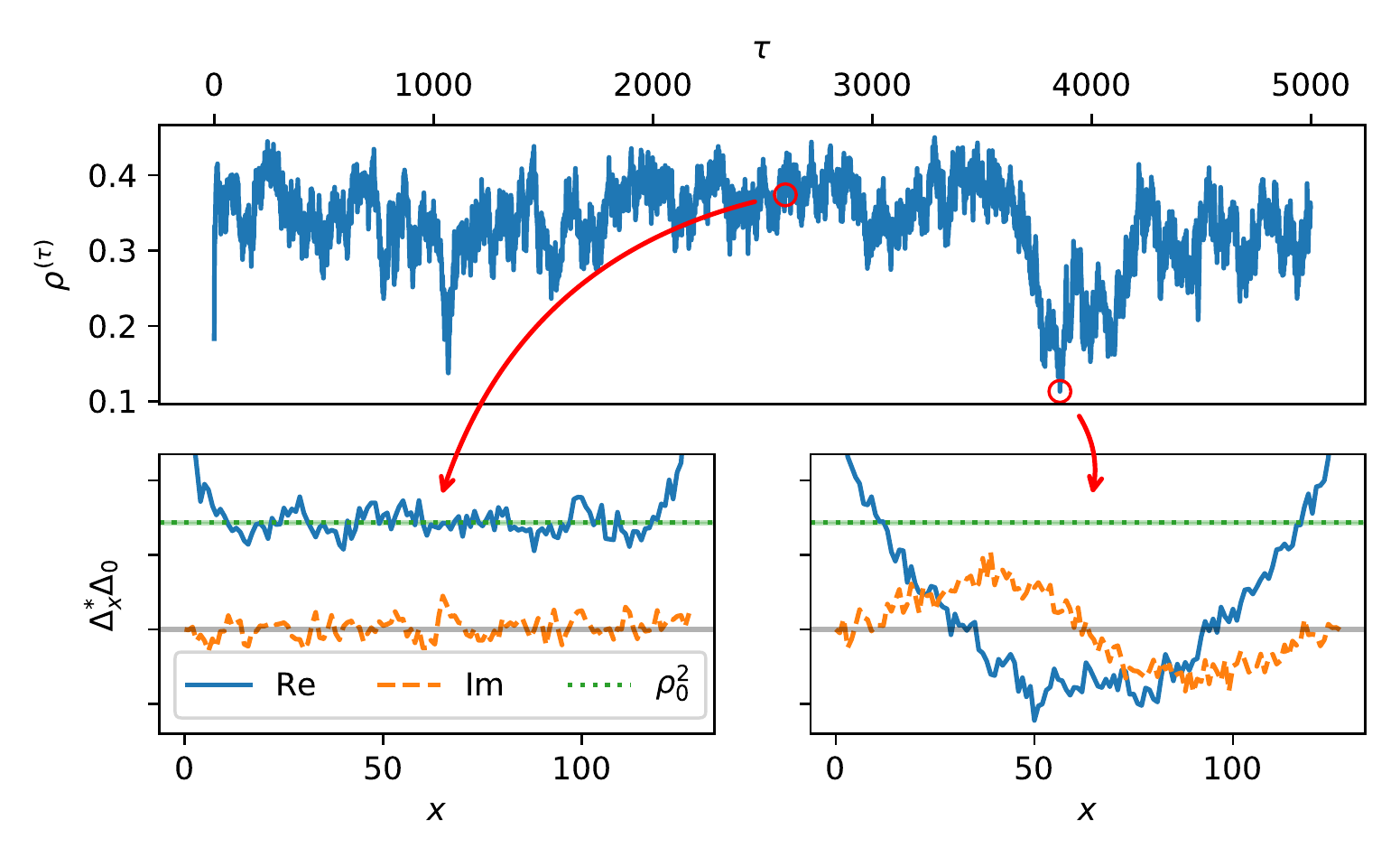}
	\caption{\label{f:scalesetting_showcase_double_peak}Monte Carlo 
		time series of $\rho^{(\tau)}$ defined in (\ref{eq:rho_scale_standardMC})
		(top) with the $\Delta_{x}^{*}\Delta_{0}$ correlator at $\tau=2604$ 
		(bottom, left) and $\tau=3856$ (bottom, right) for $\Nt=72$, $\Ns=255$,
		$g^{-2}=1.0540$ and $\mu=0$.}
\end{figure}

The diagrams for systems with constant lattice spacing
in \fref{f:pd_nf2_infinite_volume} show that the infinite-volume limit significantly shrinks the red ($C_{\min}>0$, i.e.\@ predominant homogeneous contributions) region without affecting the blue and green region of predominant inhomogeneous resp.\@ symmetric configurations.
There are three rather different phenomena at work here:
\begin{enumerate}
	\item The simplest one is just geometrical: When we enlarge the spatial volume, we can fit larger wavelengths into the finite box. For small $\mu$ the pitch of the chiral spiral 
	would exceed the box size, which means that the chiral spiral
	does not fit into the box. Such a suppression of chiral spirals 
	with large pitches disappears for larger volumes. Hence, the region of predominant homogeneous configurations must shrink in the direction of non-vanishing $\mu$.
	\item Finding a shrinking of the $C_{\min}>0$ region in the temperature direction is clear evidence against spontaneous symmetry breaking. In fact, the (qualitative) behavior of the apparent transition temperature and the condensate is well understood by a comparison of the analytically known correlation length \eqref{eq:corrBKT1b} with the box size. We can thereby clearly identify the remnant condensates that were measured as finite-size effects. 
	\item The transition from the blue ($C_{\min}<0$, i.e.\@ predominant inhomogeneities) to the green ($C_{\min}\approx 0$, i.e.\@ predominantly symmetric) regime can be easily understood as the following short-range effect: At finite temperature there are two length scales in the system (besides the finite box size), i.e.\@ the temperature-induced finite correlation length $\xi_\beta$ from \eqref{eq:corrBKT1b} and the predominant wavelength inversely proportional to $\mu$ (up to discretization due to the finite box size). Obviously, for $C_{\min}$ to be negative, the amplitude, which decays due to $\xi_\beta$, must not have dropped to (almost) vanishing values at separations where the first minimum of the oscillations occurs. Since the latter is given by $\mu$ (up to a constant factor), the transition line from blue to green signals that the temperature scale takes over as the shortest relevant scale from the chemical potential. This is not really a qualitative change. As this is independent of the much larger box size, it is not affected by the infinite-volume limit.
\end{enumerate}
An interesting, but unfortunately hard to quantify observation is the following.
While on smaller lattices (e.g.\@ $\Ns=63$) the data tend to fluctuate around only one background configuration, like a chiral spiral with a fixed winding number, larger lattices admit changing the winding number \emph{more} often as opposed to \emph{less} often. An example is depicted in Fig. \ref{f:scalesetting_showcase_double_peak}, 
which shows a time series of the modulus $\rho^{(\tau)}$ of the
spacetime average of $\bar\Delta^{(\tau)}$ defined in \eqref{eq:rho_scale_standardMC} 
of \aref{app:scale}.  
Even for vanishing chemical potential and low temperature,
the regime in the phase diagram where we set the scale and where the dominant configurations
are homogeneous, we still find several occasions on which there are dominant inhomogeneous contributions. 

For most of the MC time $\rho^{(\tau)}$ fluctuates about a constant value. During
this time the real part $C_{\sigma\sigma}$ of $C$ defined in (\ref{eq:deltastar_delta_correlator})
is almost constant and the imaginary part $C_{\sigma\pi}$ is 
small (recall that $\langle C_{\sigma\pi}\rangle=0$).
But at several MC times, e.g.\@ $\tau\approx 1100$ or  $3860$, the field $\rho^{(\tau)}$ drops
and the real and imaginary parts of $C(x)$ show the profiles
typical for a chiral spiral.
While the lower left 
plot in Fig. \ref{f:scalesetting_showcase_double_peak} 
is representative for most of the configurations, 
the sudden drops in the time series are strongly 
correlated with the appearance of inhomogeneous configurations 
as seen in the lower right panel.
That this is only seen on large lattices is counterintuitive at first since autocorrelation times usually increase with the system size and it is also the opposite of what was observed for the $\Z_{2}$ GN model during the work on \cite{LPW20, LPW20_1}.
However, the fact that considerable phase fluctuations on large scales are allowed is the analytically predicted mechanism for avoiding spontaneous symmetry breaking, see \sref{sec:analytical}. From that perspective, it supports the analytical claims. Obviously, \fref{f:scalesetting_showcase_double_peak} showcases a large autocorrelation time $\tau$, which, however, is under control due to good statistics of the order of $\gtrsim20\tau$.

The phase diagrams for systems with approximately constant physical
volume and successively smaller lattice spacing are shown in \fref{f:pd_nf2_continuum}.
As can be seen, we find inhomogeneities\footnote{As discussed previously, these are probably all short- and intermediate ranged, although their range does exceed the finite box size at low temperatures.} for all our lattice spacings and the results are consistent with their existence in the continuum limit.
Unfortunately, setting the scale in a partially conformal system is a very subtle issue as the dominant fluctuations have no scale at all (at zero temperature).
Since the details of this scale setting procedure are highly non-trivial 
(see \aref{app:scale}), we must leave a more detailed
analysis to a future publication.

As is discussed in detail in \aref{app:autocorrelation}, our simulations suffer from large autocorrelations. For a large region in parameter space on all geometries these autocorrelations are under control due to sufficient statistics. However, close to the critical line at $T=0$ autocorrelation times diverge and the shown data should only be regarded as qualitative in the sense that they surely capture the important phenomena found in large but finite regions of space but might be off quantitatively due to autocorrelation-related suppression of subdominant local minima. However, the discussion of \aref{app:autocorrelation} makes it clear that these will not affect our conclusions.

We conclude that we observe
inhomogeneous structures in the cGN model with only $2$ flavors
-- similarly as in the large-$\Nf$ model. The notable difference is that -- in accordance with pertinent no-go theorems -- these are incoherent on sufficiently large scales thereby hindering spontaneous symmetry breaking.
A comparable study of the $\Z_2$ GN model with $8$ flavors in \cite{LPW20}
led to a similar conclusion: inhomogeneous structures persist in the
infinite volume limit. 
We cannot say for certain whether this
remarkable feature survives the continuum limit of the
cGN lattice models.

\subsection{Decay properties of \texorpdfstring{$C_\mathrm{4F}$}{C}}

\begin{figure}[h]
	\includegraphics[scale=0.5]{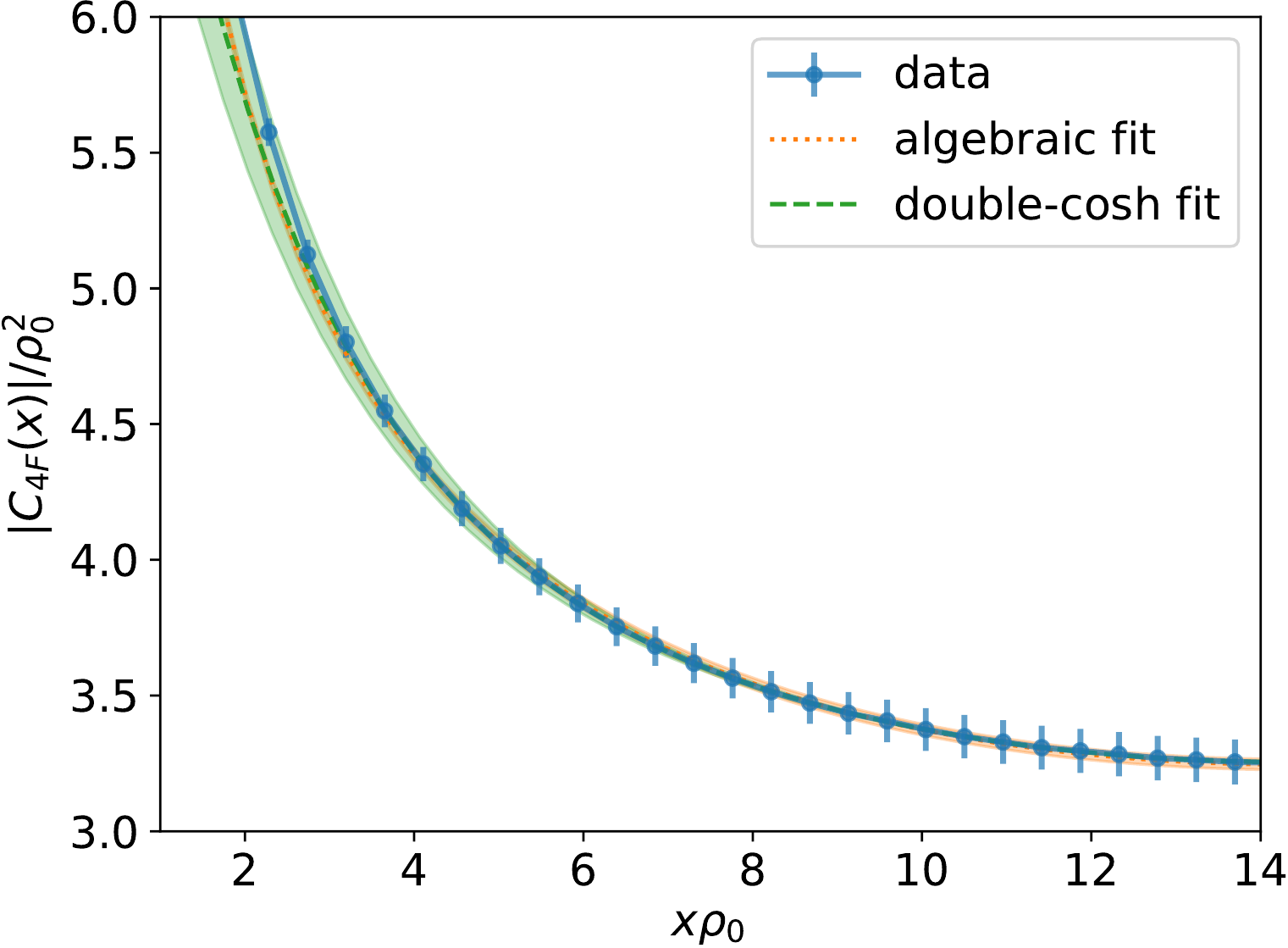}
	\caption{Modulus of the connected four-point function $C_\mathrm{4F}$ (from \eqref{eq:wi9}) with algebraic and exponential fits whose 
	fit parameters are given in (\ref{fitpar1}) and (\ref{fitpar2}).}
	\label{f:witten}
\end{figure}

We analyzed the decay properties of $C_\mathrm{4F}(x)$ as given by \eqref{eq:wi9} 
on a $72\times63$ lattice with $a\rho_0\approx0.46$, i.e.\@ at a temperature $T\approx0.03\rho_{0}$. In order to study its
infrared behavior we computed the connected correlation function.
Motivated by the asymptotic forms (\ref{asywitten}) predicted
by the low-energy effective action
we fit the data points via a (symmetrized) algebraic function
\begin{equation}
	C_{\mathrm{4F}}(x) = \frac{\alpha}{x^\beta}+\frac{\alpha}{(L-x)^\beta}\;,
\end{equation} 
as well as a double-cosh ansatz,
\begin{equation}
	C_{\mathrm{4F}}(x) = \sum_{i=1}^2\gamma_i\cosh\left[m_i\left(x-\frac{L}{2}\right)\right]\;,
\end{equation}
and show the results in \fref{f:witten}.

\noindent The fit parameters for a power-law decay are
\begin{equation}
\frac{\alpha}{\rho_0^2}=6.52\pm0.02\;,\quad \beta=0.521\pm0.001
\label{fitpar1}
\end{equation}
and for an exponential decay we find
\begin{align}
\begin{aligned}
\frac{\gamma_1}{\rho_0^2}=3.2515\pm4\cdot10^{-4}\;,\quad &\frac{m_1}{\rho_0}=(5.76\pm0.03)\cdot10^{-2}\;,\\
\frac{\gamma_2}{\rho_0^2}=(4.4\pm0.3)\cdot10^{-3}\;,\quad &\frac{m_2}{\rho_0}=0.531\pm0.006\;.
\label{fitpar2}
\end{aligned}
\end{align}

These results confirm similar findings obtained for the $\Z_2$ GN model, namely that it is very difficult
to distinguish between power-law and exponential decays on the lattices
with $N_s=63$, which was also to be expected following the previous discussion 
and \cite{BH94}. 
However, from the perspective of our analytical discussion, where we predicted a massive phase for any $T>0$ with the mass vanishing in the limit $T\to 0$, there is a very well-motivated explanation. \eqref{eq:corrBKT1b} predicts
\begin{align}
	m_{1}=1/\xi_\beta\approx 2\cdot 10^{-2}\rho_0\;,
\end{align}
which is reasonably close to the fitted value (remember that we expanded in $\mathcal{O}(1/\Nf)\sim 50\%$ for $\Nf=2$) and explains its seemingly unnaturally small magnitude.

On the other hand, we find that
the value $\beta\approx0.52$ is only marginally larger than the theoretical zero-temperature
prediction of $\beta=0.5$ for two flavors in \eqref{decaywitten}.
This result -- although not precisely a proof -- is in astonishing agreement with the analytical prediction coming from an expansion around $\Nf=\infty\gg 2$ and furthermore beautifully reveals how the massive phase more and more approximates the conformal behavior at zero temperature by the unexpectedly large mass ratio $m_2/m_1 \approx 10$.

\subsection{The phase field \texorpdfstring{$\theta$}{theta}}

In this section we analyze $\left\langle \theta(x) \right\rangle$. 
This discussion should be regarded as complementary to the previous analysis of the correlators in the sense that we now use a quantity 
directly related to the fields.
It will further substantiate our previous findings.

\begin{figure}[h]
	\includegraphics[scale=0.5]{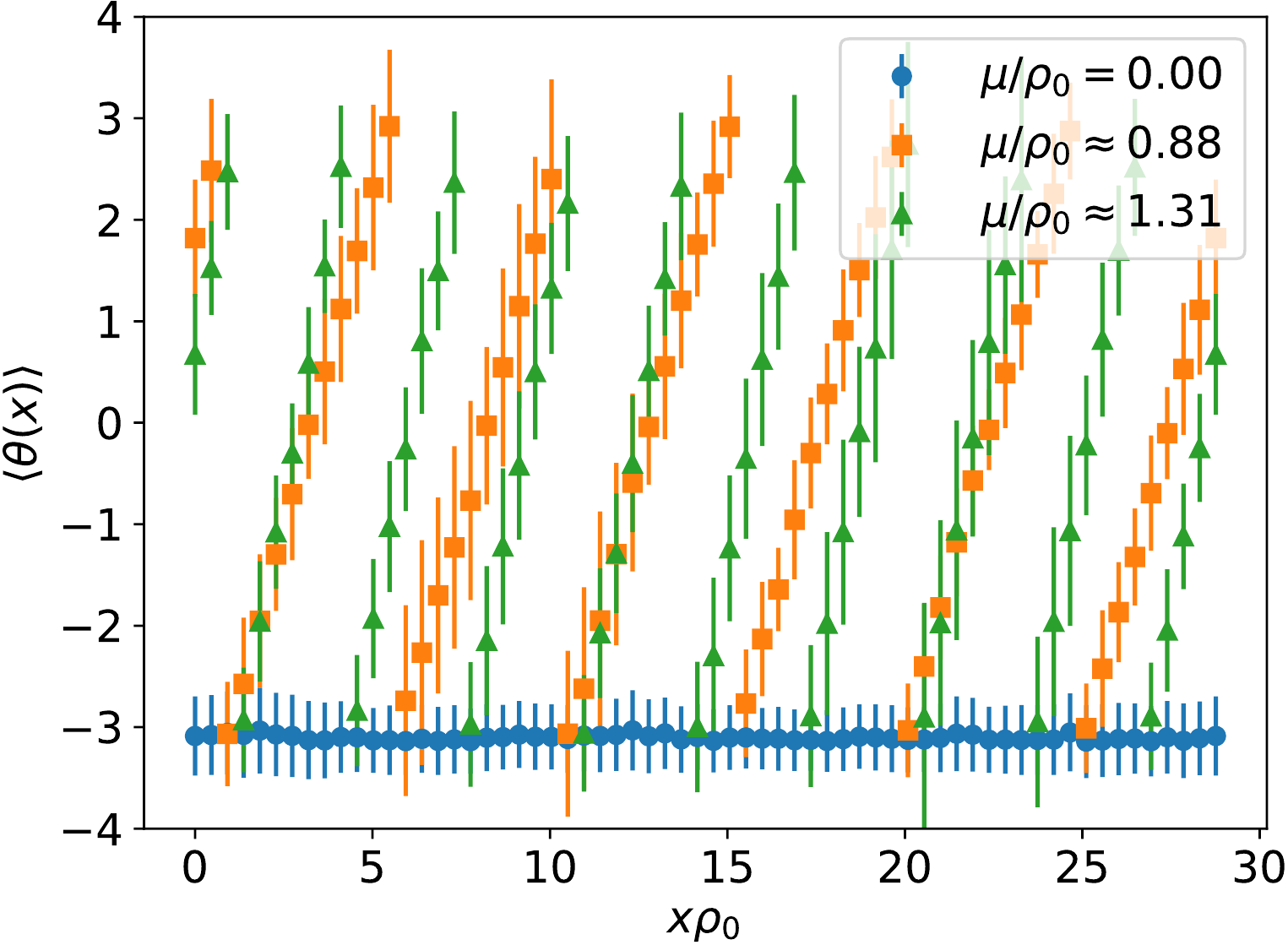}
	\caption{Angle of the averaged scalar field, $\langle\theta\rangle=\arg(\langle\bar{\Delta}\rangle)$, 
		calculated from time-averaged fields $\bar\sigma$ 
		and $\bar\pi$, as a function of the spatial coordinate for different $\mu$ 
		at $N_t=72$.}
	\label{f:theta}
\end{figure}

We show the $x$ dependence of the average $\langle\theta(x)\rangle$, 
as defined in \eqref{eq:avtheta}, on a $72\times63$ lattice 
for $a\rho_0\approx0.46$ and for three values of the chemical 
potential in Fig. \ref{f:theta}.
 For vanishing $\mu$ the argument of the averaged complex
condensate field $\Delta$ is constant,
which means that the latter does not wind.
For the intermediate value $\mu\approx0.88\rho_0$ the 
average angle is an almost linear function
of $x$ and the complex condensate winds $6$ times when one moves
along the spatial direction.
When one further increases the chemical potential to $\mu\approx1.31\rho_0$, 
the slope of the (almost) linear mapping $x\mapsto\langle \theta(x)\rangle$ increases and the
condensate winds $9$ times.

We see that the winding number of the chiral spiral around the spatial 
direction increases with increasing $\mu$. In fact, the winding number
extracted from the averaged field $\langle \theta(x)\rangle$ 
perfectly agrees with the dominant winding number defined in \eqref{eq:nmax}
and depicted in \fref{f:nf2_nmax}. To summarize: at low temperature $\langle \theta(x)\rangle$ calculated from (\ref{eq:avtheta}) is almost a
linear function of $x$ with a slope proportional to $\mu$. In agreement with
the analysis based on the dominant wave number we detect a chiral spiral
with a winding number proportional to $\mu$.

	\section{Conclusions}\label{sec:conclusions}
	In the present work we studied the $(1+1)$-dimensional chiral Gross-Neveu model
with chiral SLAC fermions and exact axial U$_A$(1) symmetry on
the lattice.
Our two main results are summarized in the following.

First, we have strong and multiple evidence that the analytical prediction
from an expansion in $1/\Nf$ well describes the
qualitative features of the cGN model with $2$ flavors. 
As expected we see no spontaneous symmetry breaking with 
long-range order in the strict
sense, and our results
suggest that at $T=0$ there is a BKT phase with quasi-long-range order. For
example, the low-temperature regime, where we have signals of (in)homogeneous
ordering over the whole lattice, is well explained by the analytically
predicted correlation length exceeding the finite box size and it shrinks
consistently in the thermodynamic limit. Additionally, the decay properties of
pertinent correlation functions are well fitted by the analytically predicted
ans{\"a}tze with reasonable parameter values.

The latter suggests that for $T=0$ fluctuations of the phase
field $\theta$ on all scales exist and are responsible for the
restoration of the axial symmetry. We demonstrated that $\theta$
is uniformly distributed on unit circles in complex field space
and that large system sizes allow for long-range phase fluctuations
strong enough to change the winding number.
This behavior is predicted by 
the effective low-energy theory for $\theta$ which has been
taken from \cite{And05} and extended to $\mu\neq 0$.

Despite this, our second finding is that, rather unexpectedly,
our simulations at finite temperature and density 
reveal that the cGN model with only $\Nf=2$ flavors resembles 
the analytic large-$\Nf$ solution in many ways.
The chiral spirals are still seen in the dominant configurations
and their winding numbers increase linearly with
the chemical potential.
The only qualitative difference at low temperatures is that these structures are only coherent in finite but -- depending on the temperature -- potentially very large regions of space.
Instead of a temperature-driven phase transition at intermediate temperatures, we found a competition of the two important scales in the system, viz. the temperature-induced finite correlation length and the density-induced wavelength. So, the question whether or not oscillating behavior was observed (on potentially short scales) can be answered only from comparison of the wavelength with the correlation length. Or, put differently, it is very likely that oscillating behavior can be found for any temperature and non-vanishing chemical potential as long as the wavelength is shorter than the correlation length of the system. This is qualitatively different from the large-$\Nf$ behavior where there is a strict critical temperature above which no oscillation can be observed. 

We have verified these results mainly via the correlator $C$
in (\ref{eq:deltastar_delta_correlator}) and by analyzing the phase of the averaged field $\Delta$, defined in (\ref{eq:avtheta}).
We generated many ensembles for the control parameters $T$
and $\mu$ on grids with up to $192$ points.
To quantify finite-size and discretization effects the simulations
were repeated on lattices with $63,127$ and $255$ points in the spatial
direction. While we have good signals for the behavior in the thermodynamic limit, whether the inhomogeneities
remain after the continuum limit has been taken is less clear.
With the chosen scale setting, which is a subtle issue in 
a theory with quasi-long-range order, we observe that inhomogeneities remain in the limit $a\rho_0\to 0$.
We hope to gain a more thorough understanding of this limit in the future.

Although we found strong evidence that consistently supports the analytical predictions, our method of MC simulations will never be able to prove this in a rigorous sense. Therefore, it would
be interesting to compare our findings with results from other methods,
for example the functional renormalization group. It would
be valuable to continue the study of the $(1+1)$-dimensional 
Gross-Neveu-Yukawa model in \cite{Stoll:2021ori} to related systems
in finite volumes and inhomogeneous background fields.

The mechanism of how the cGN-model realizes the U$_A(1)$ symmetry
is similar to the flattening of the constraint effective potential 
for a spacetime-averaged order parameter $\bar{\Delta}$  \cite{ORaifeartaigh:1986axd}.
For example, in the Ising model at low temperature, if 
we impose that the spatially averaged spin
vanishes in the sum over spin configurations, then in a typical
configuration we observe large regions with spin up and large regions 
with spin down. Despite the surface energy stored in the walls separating 
the "up" and "down" regions,
this is the energetically preferred way of fulfilling the external constraint.

Models with a continuous symmetry react differently to the constraints. 
For example, in the $3$-dimensional O(2) model with a Mexican hat potential
for a complex scalar field $\Delta$ the constraint 
$\vert\bar\Delta\vert<\langle \vert\Delta\vert\rangle$
is met by inhomogeneous spin-wave-like configurations with $\vert\Delta(\vx)\vert\approx \langle \vert\Delta\vert\rangle$
\cite{endrodi}. These
configurations resemble the chiral spiral in the cGN model, 
for which the modulus of $\bar\Delta$ can
be much smaller than $\langle \vert\Delta\vert\rangle$. 
In the $2$-dimensional cGN model the constraint $\bar\Delta\approx 0$ 
is not imposed by hand but by general theorems which ensure that $\langle \bar\Delta\rangle=0$.
In a typical configuration the modulus of $\Delta(\vx)$ is near the minimum
$\bar\rho$ of the effective potential -- in order to minimize the bulk
energy -- but the real and imaginary parts $\sigma$ and $\pi$
have vanishing expectation values caused by large phase fluctuations about
the relevant chiral spiral. The main difference between the 3-dimensional
O(2) model and the 2-dimensional cGN model is that in the former
model the wavelength of the inhomogeneity is given by the box size
\cite{endrodi} and in the latter by the inverse chemical potential.

In \cite{PTV20} it has been emphasized
that the occurrence of correlation functions 
exhibiting damped oscillations in the spatial directions 
is directly related to particular features of the 
dispersion relations. The associated quantum spin liquid
behavior, which we also spotted in the $2$-flavor cGN model, may thus 
be observed in a larger class of field theories.

After publishing the initial draft of our manuscript a similar study of the (1+1)-dimensional cGN model using the naive fermion discretization was published in \cite{HN21p}. Its results are in qualitative agreement with ours.

	\begin{acknowledgments}
		We thank Björn Wellegehausen for providing the code base used in the present 
		work and for fruitful discussions. In addition, we thank
		Laurin Pannullo, Marc Wagner and Marc Winstel for many discussions on
		four-Fermi theories and the collaboration during a previous work
		on the $\Z_2$ GN models.
		This work has been funded by the Deutsche Forschungsgemeinschaft (DFG) under 
		Grant No. 406116891 within the Research Training Group RTG 2522/1. 
		The simulations were performed on resources of the Friedrich Schiller 
		University in Jena supported in part by the DFG Grants INST 275/334-1 FUGG and 
		INST 275/363-1 FUGG, as well as for some pioneer runs during the work on \cite{LPW20,LPW20_1} on the GOETHE-HLR high-performance computer
		of the Frankfurt University.
	\end{acknowledgments}
	
	\appendix
	\section{Details of scale setting}
	\label{app:scale}
	\begin{figure*}
		\begin{subfigure}[t]{.32\linewidth}
	\includegraphics[scale=0.39]{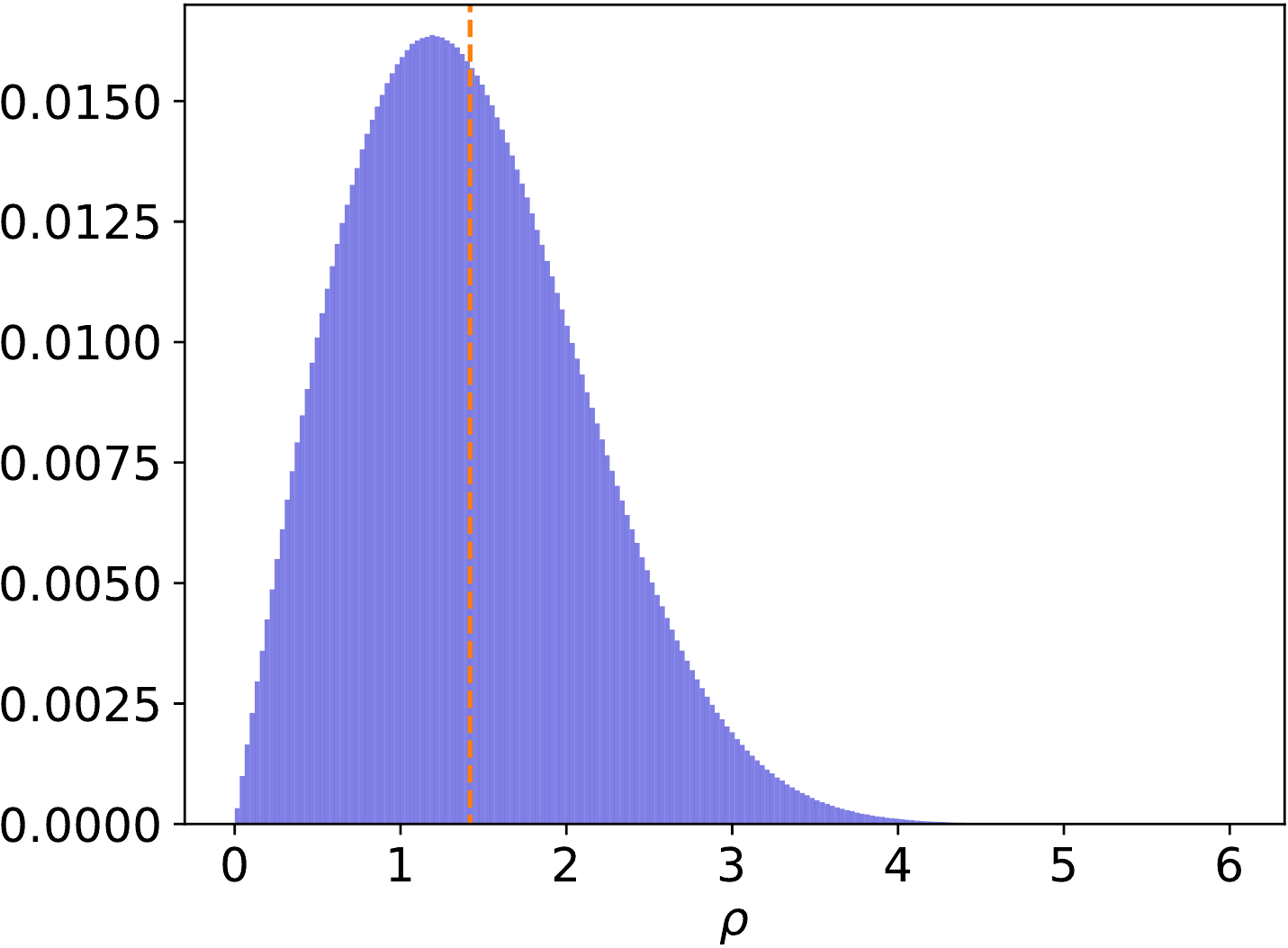}
	\subcaption{Distribution of which \eqref{eq:rho_scale_fieldtheory} is the average.}
	\label{f:scalesetting_showcase_1} 
		\end{subfigure}
		\begin{subfigure}[t]{.32\linewidth}
	\hspace{0.3cm}
	\includegraphics[scale=0.39]{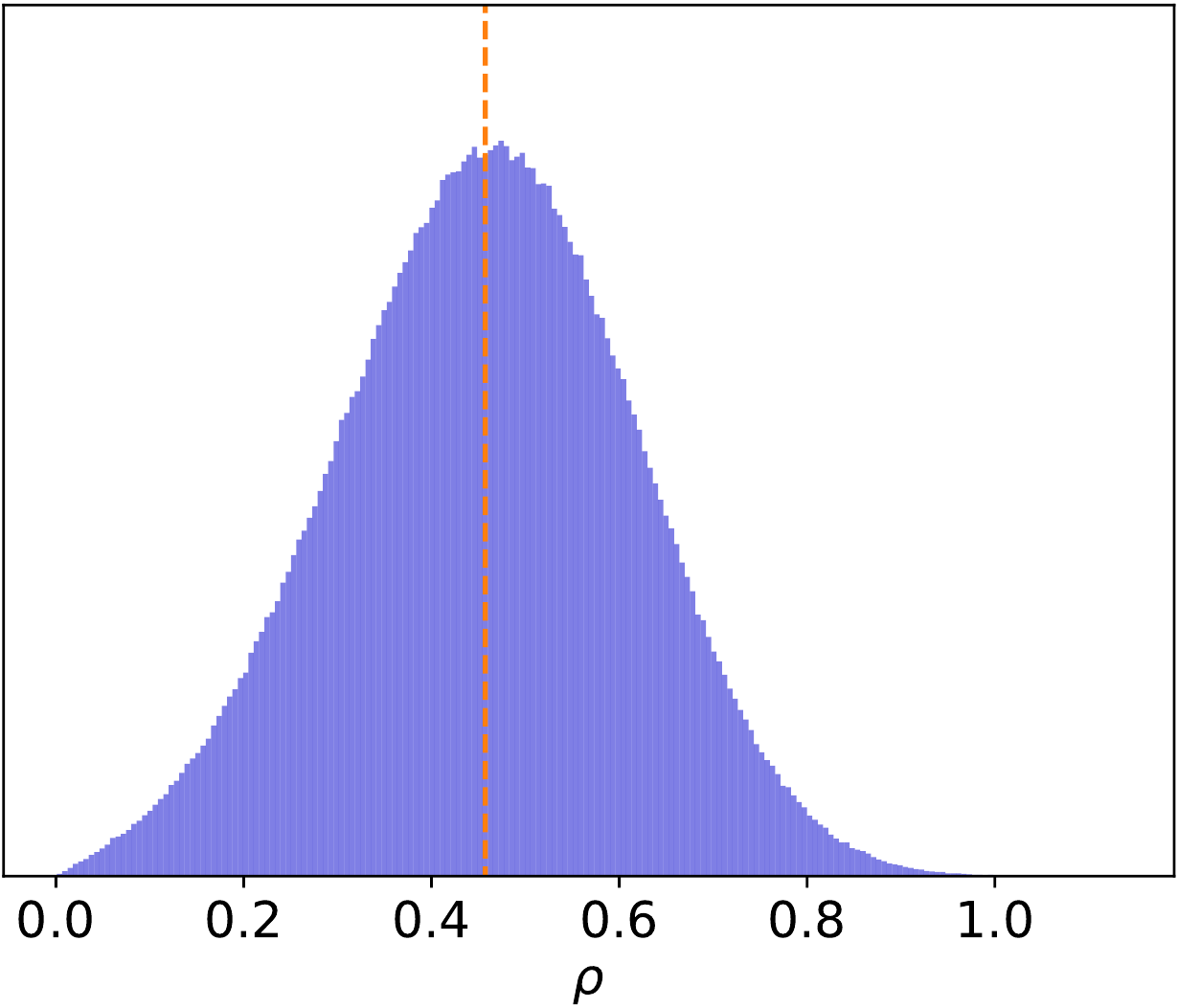}
	\subcaption{\label{f:scalesetting_showcase_2} Distribution of which \eqref{eq:rho_scale_final} is the average.}
		\end{subfigure}
		\begin{subfigure}[t]{.32\linewidth}
	\includegraphics[scale=0.39]{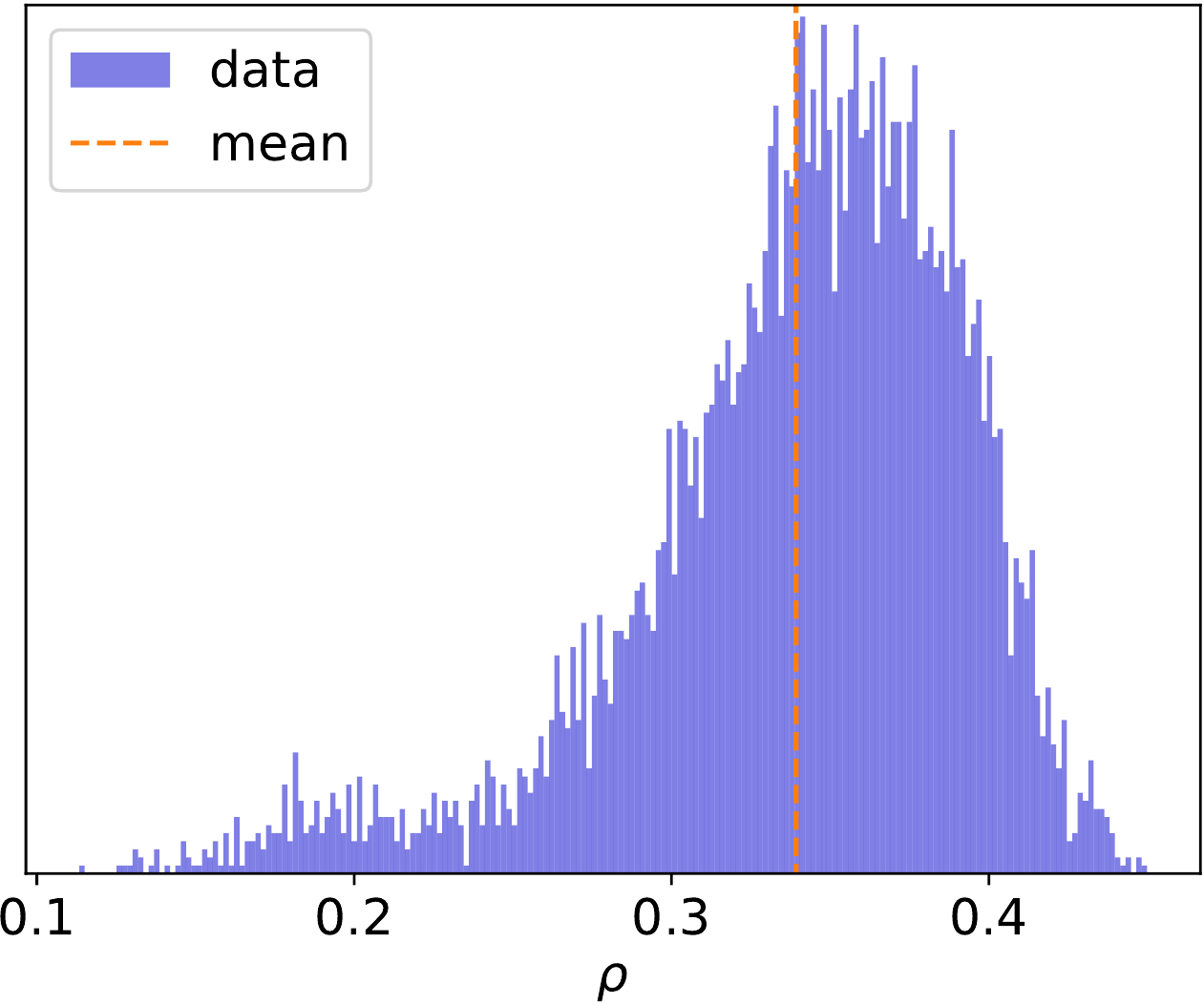}
	\subcaption{\label{f:scalesetting_showcase_3} Distribution of which \eqref{eq:rho_scale_standardMC} is the average.}
		\end{subfigure}
	\caption{\label{f:scalesetting_showcase}Histogram data of various estimators of $\left\langle \rho \right\rangle$ for $\Ns=255$, $\Nt=72$, $\mu=0$ and $g^{-2}=1.0540$. The vertical axes show the relative number of hits per bin, with 200 bins used for each histogram. Note that all values of $\rho$ shown here are raw numerical data not in physical units.}
\end{figure*}

For an easy comparison of our results with the analytic large-$\Nf$ 
solution we use $\rho_{0} = \left\langle \rho \right\rangle_{T=\mu=0}$
to set the scale. Unfortunately, it is difficult to obtain
an accurate estimate for $\langle\rho\rangle$ in our simulations. 
In this appendix, we first explain the (statistical) problems 
with direct approaches to measure $\left\langle \rho \right\rangle$ 
and afterwards present our solution. 

From a field theory perspective, the direct lattice estimator 
for $\left\langle \rho \right\rangle$ would be $\rho_{t,x}$ 
for any (fixed) point $(t,x)$ on the lattice. Now, $\rho_{t,x}$ 
should be homogeneous up to fluctuations and, hence, one can 
improve the statistics by combining the data from all 
estimators $\rho_{t,x}$ for all lattice points. 
Example data for this estimator can be seen in \fref{f:scalesetting_showcase_1}.

The final estimate for $\left\langle \rho \right\rangle$ would then 
read
\begin{equation}
	\label{eq:rho_scale_fieldtheory}
	\left\langle \rho \right\rangle \approx \left(\mathrm{mean}_{\tau}\circ\mathrm{mean}_{x}\circ\mathrm{mean}_{t}\circ \mathrm{abs}\right)
	\Delta_{t,x}^{(\tau)}\;,
\end{equation}
where $\tau$ is the MC time, $\Delta_{t,x}^{(\tau)}$ the field value at site
$(t,x)$ of the $\tau$'th configuration and $\mathrm{mean}_{\#}$ 
means averaging with respect to the respective subscript. In order to actually 
show the distribution from which the final estimates are calculated,
we present (here and in the following) the histograms one obtains by stripping the means
\emph{after} the absolute values have been taken.

The histogram of the straightforward estimator shown in \fref{f:scalesetting_showcase_1} 
is dominated by 
its broad variance (as is expected for a local estimator). 
More importantly, since the field $\theta$ is quasi-long-range,
it requires many sweeps through the lattice to obtain a $\theta$-independent
distribution of $\Delta$ like the ones depicted in \fref{fig:2dhistogram}. 
In fact, a typical configuration in the simulations is not distributed symmetrically 
around the origin but rather around some finite value $\Delta_0$.
The center of the configurations moves slowly (in Monte Carlo time) around the origin in field space.
For this reason, taking the modulus right in the beginning 
leads to a significant bias towards larger 
values in the estimator (\ref{eq:rho_scale_fieldtheory}).

The broad variance mentioned above is a known statistical phenomenon in MC simulations and is 
usually cured by averaging over the spacetime lattice \emph{before} 
taking the absolute value, schematically
\begin{equation}
	\label{eq:rho_scale_standardMC}
	\left\langle \rho \right\rangle \approx \left(\mathrm{mean}_{\tau}\circ \mathrm{abs}\circ\mathrm{mean}_{x}\circ\mathrm{mean}_{t}\right)
	\Delta_{t,x}^{(\tau)}\;.
\end{equation}
This sharpens the distribution but 
is less well motivated from a field theory perspective. The
choice (\ref{eq:rho_scale_standardMC}) can be justified 
if there is spontaneous symmetry breaking
and a small trigger is sufficient to align the values of the 
field on the lattice sites. In this case 
the absolute value does not change the result if we take
the limits in the correct order, i.e.\@ the spatial volume to infinity before 
removing the trigger. In the symmetric phase, on the other hand, 
already the spatial average should vanish in 
the thermodynamic limit and again taking the absolute value does not make a difference. 
Example distributions of this estimator are shown in 
\fref{f:scalesetting_showcase_3}. 
Note the different scales on the $x$ axes.

It may come as a surprise that 
there is a second peak visible that distorts the mean 
of this distribution. This is due to the fact that at any non-zero 
temperature there are contributions from inhomogeneous configurations, which average 
out over the lattice to a very good approximation, see also \fref{f:scalesetting_showcase_double_peak}. While for these data 
the distortion might be mild, we are not willing to take the risk of 
severely underestimating the observable for scale setting.

In the present work, what is even more problematic is that long-range 
(quasi-periodic) inhomogeneities must 
not be averaged over the spatial direction before taking absolute
values. But, since we have to improve statistics as much as possible
we will compromise by using
\begin{align}
	\label{eq:rho_scale_final}
	\left\langle \rho \right\rangle \approx \left(\mathrm{mean}_{\tau}\circ\mathrm{mean}_{x}\circ \mathrm{abs}\circ\mathrm{mean}_{t}\right)
	\Delta_{t,x}^{(\tau)}\;,
\end{align}
where we, similarly as in the spatial correlation functions (\ref{eq:spatial_correlators}),
first average over time.

As \fref{f:scalesetting_showcase_2} indicates, this yields acceptable statistics 
while only using the assumption of temporal homogeneity which is a 
feature of all large-$\Nf$ results we know of and was checked to be 
valid in our MC data, see, for example, \fref{f:sigma_configuration}. One should note that this procedure 
does not work in the high-temperature regime as the distribution in this case
approaches that of \eqref{eq:rho_scale_fieldtheory}.

In future works other scale settings could be used and 
the corresponding results should be compared
with those obtained in the present work. For example, the mass of
the field $\rho(t,x)$ may serve as an energy scale. The drawback of 
choosing a scale different from the minimum of the effective potential
$U_\mathrm{eff}(\rho)$ (at zero temperature and density) is
that it is less straightforward to relate to the analytic 
results for large $\Nf$. In the large-$\Nf$ limit the
field $\rho$ becomes infinitely heavy.

	\section{Autocorrelation analysis}
	\label{app:autocorrelation}
	During our simulations, we had to tackle severe autocorrelations similar to those described in \cite{LPW20_1}. In this appendix we summarize our extensive analysis of autocorrelation functions (ACFs) of various lattice estimators and provide details on how we arrive at the conclusion that our qualitative statements are robust despite large autocorrelations for certain parameter regions.

\subsection{Identifying autocorrelation scales in an example}
To facilitate such a discussion it is useful to visualize the topography of the effective action of the theory. In the infinite volume and -- less important for this argument -- continuum case, \cite{BDT09} found the general form of the saddle points of the effective action. The spatial profiles of the order parameter $\Delta$ are given as a continuous family with four parameters related to overall scale, amplitude and phase variations, and a phase offset which is tightly related to translations. The finite volume we work in subjects these self-consistent solutions to the imposed boundary conditions such that for some of these parameters only a discrete subset of allowed solutions yields saddle points in finite volume. This entails a ragged landscape of valleys with local minima of the effective action that are separated by ridges stemming from the finite-volume effects and melting in the infinite-volume limit. From the analytical results, we expect chiral--spiral-like local minima (including the degenerate one, i.e.\@ the constant order parameter) to be most important and our simulations confirm this expectation.

The above discussion immediately suggests that there are three qualitatively different kinds of autocorrelations: Sampled configurations will typically tend to fluctuate around one local minimum correlated within this valley on a (MC-time) scale $\tau_{\mathrm{fluct}}$. During this process the reference chiral spiral will rotate the overall phase offset on a time scale $\tau_{\mathrm{U}(1)}$ which, for non-degenerate chiral spirals, is equivalent to translating this spiral. Eventually, the algorithm will climb (or tunnel through) a ridge and arrive in another valley on a time scale $\tau_{\kmax}$.

From these three time scales, $\tau_{\mathrm{U}(1)}$ is of minor importance to us because we carefully crafted all of our observables to respect the $\mathrm{U}(1)$ (and closely related translational) symmetry. From the notable exceptions \fref{fig:2dhistogram} and \fref{f:theta}, however, we learned that it is quite sizable but clearly under control as the almost-perfect circles of \fref{fig:2dhistogram} illustrate.

\begin{figure}[h]
	\includegraphics[width=0.9\linewidth]{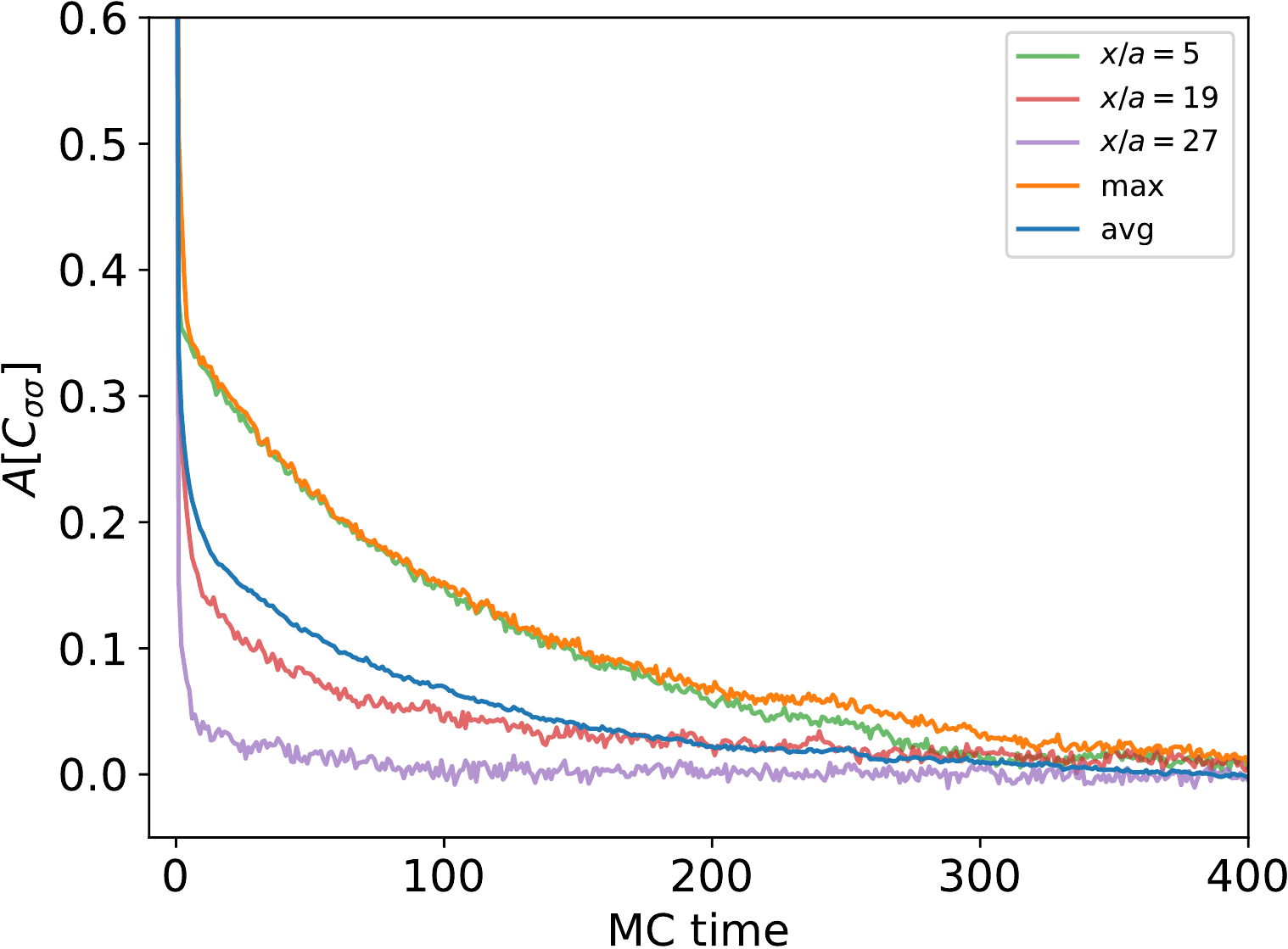}
	\caption{\label{f:spatboscorr_acf}Autocorrelation functions for $C_{\sigma\sigma}(x)$ for three different $x$, a maximum over all $x$ and an average over all $x$ for $\Nf=2$, $\Ns=63$, $T/\rho_0\approx0.09$, $\mu/\rho_0\approx0.61$, $a\rho_0\approx0.46$.}
\end{figure}

The other autocorrelation scales can be clearly distinguished in \fref{f:spatboscorr_acf}. For one exemplary parameter set, the figure shows ACFs of $C_{\sigma\sigma}(x)$ for some randomly chosen lattice points $x$ as well as the average and the (local in MC-time separation) maximal autocorrelations obtained over all lattice points. The latter rather unconventional quantity can be considered as a worst-case scenario for autocorrelations in $C_{\sigma\sigma}$. 

All of these are well described by an ansatz
\begin{align*}
	A[C_{\sigma\sigma}](t) = be^{-t/\tau_1}+(1-b)e^{-t/\tau_2}\;,
\end{align*}
where $b$, $\tau_{1}$ and $\tau_{2}$ are free parameters. While the detailed numbers obviously depend on the data set chosen for fitting, the orders of magnitudes are consistent (cf.~\tref{t:autocorr}).

\begin{table}[h]
	\caption{\label{t:autocorr} Fit parameters $b$, $\tau_1$ and $\tau_2$ for the autocorrelation function of $C_{\sigma\sigma}(x)$ for some values of $x$ as well as an average and a maximum over all $x$ (cf. \fref{f:spatboscorr_acf}).}
	\newcommand{\ncol}{4}
	\renewcommand{\tabcolsep}{4pt}
	\begin{tabular}{ccccc}
		\hline \hline
		& $b$ & $\tau_1$ & $\tau_2$\\
		\hline
		\rule[-1ex]{0pt}{5ex}
		$x/a=5$ & 0.65 & 0.3 & 111.5\\
		\hline
		\rule[-1ex]{0pt}{5ex}
		$x/a=19$ & 0.86 & 0.8 & 101.5\\
		\hline
		\rule[-1ex]{0pt}{5ex}
		$x/a=27$ & 0.94 & 0.5 & 32.8\\
		\hline
		\rule[-1ex]{0pt}{5ex}
		avg & 0.79 & 0.6 & 87.0\\
		\hline
		\rule[-1ex]{0pt}{5ex}
		max & 0.66 & 0.9 & 125.5\\
		\hline
		\hline
	\end{tabular}
\end{table}

\begin{figure*}
	\begin{subfigure}[t]{.49\linewidth}
		\includegraphics[width=\linewidth]{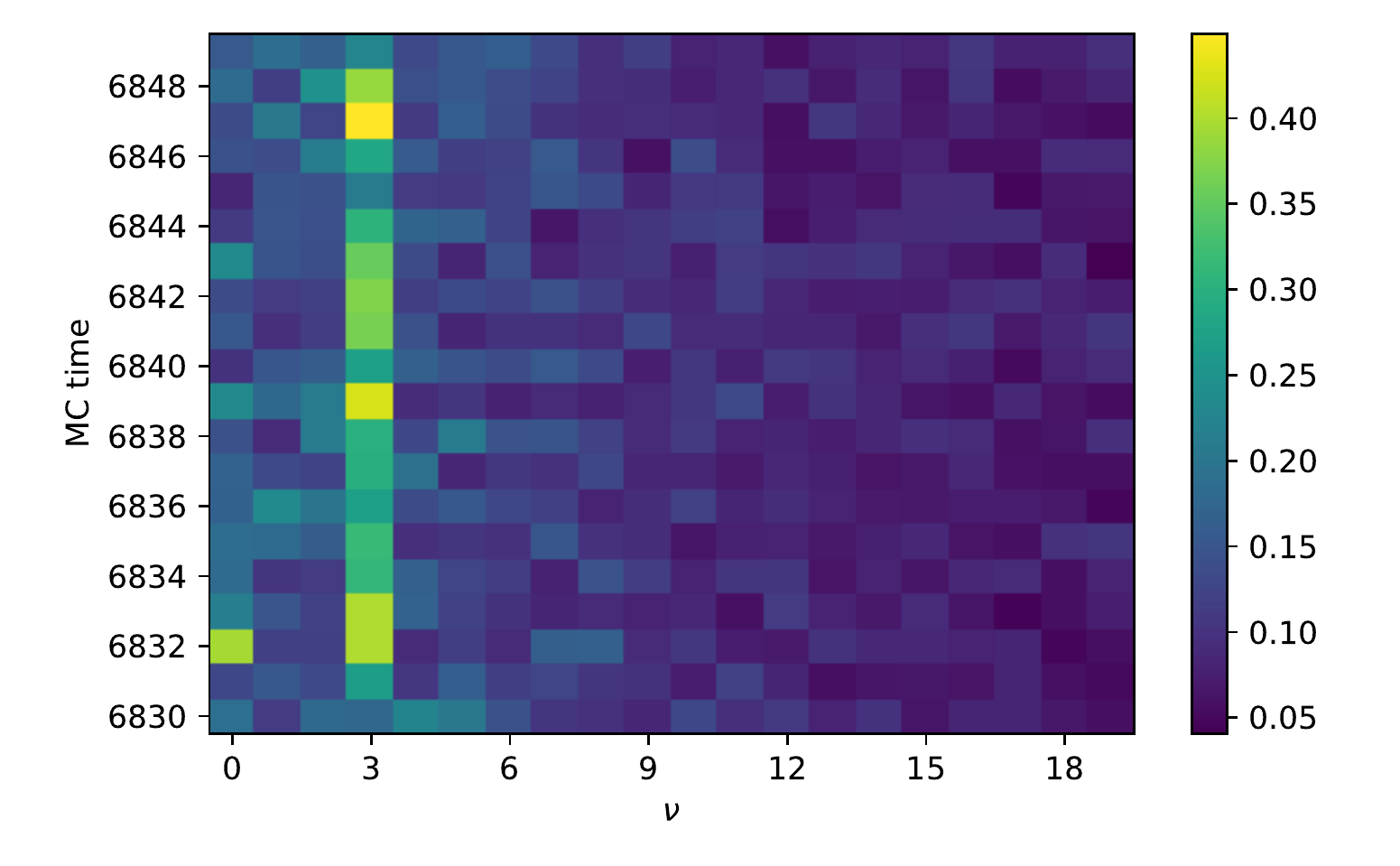}
		\caption{\label{f:spatboscorr_history_zoom}Short MC time scales.}
	\end{subfigure}
	\hfill
	\begin{subfigure}[t]{.49\linewidth}
		\includegraphics[width=\linewidth]{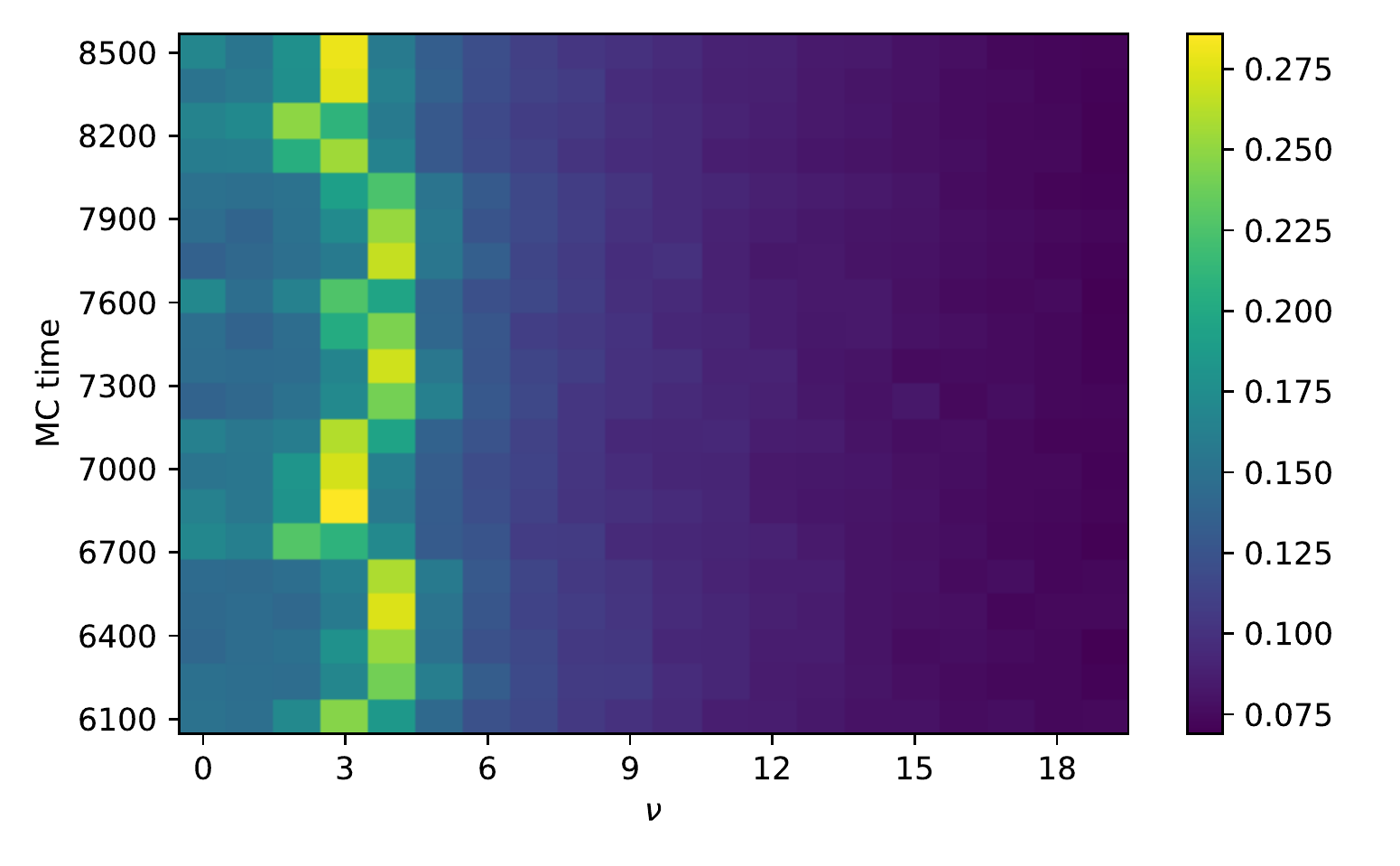}
		\caption{\label{f:spatboscorr_history_binned}Long MC time scales. The data was averaged over bins of size $\tau_{\kmax}=\tau_{2}\approx126$.}
	\end{subfigure}
	\caption{\label{f:spatboscorr_history}MC timelines of the Fourier spectra of $C_{\sigma\sigma}$ for the same parameters as \fref{f:spatboscorr_acf}.}
\end{figure*}

In order to associate the two numerical values $\tau_{1}$ and $\tau_{2}$ with the mechanisms described above, we consider representative time evolutions on each time scale in \fref{f:spatboscorr_history}. \fref{f:spatboscorr_history_zoom} shows the Fourier spectrum of $C_{\sigma\sigma}$ over $20\times \tau_{1}$ MC configurations where we conservatively use $\tau_{1}\approx 1$. It is clearly seen that there is a constant peak at $\nmax = 3$, while the MC evolution produces small fluctuations around this reference configuration. We conclude that the small time scale for this parameter set is generated from fluctuations around one local minimum, i.e.\@ $\tau_{\mathrm{fluct}}=\tau_{1}$.

To probe the larger time scale, we show a MC-time window of $20\times \tau_{2}$ configurations in \fref{f:spatboscorr_history_binned}, where we conservatively estimated $\tau_{2}\approx 126$. For visual clarity, we averaged blocks of $\tau_{2}$ MC configurations which should be thought of as a coarse graining integrating out the high frequencies similar to an RG transformation. On this scale, the correlator spectra are smooth (due to the coarse graining) and sharp peaks and the MC evolution produces jumps in the dominant frequency $\kmax$. This finding relates the long time scale to $\tau_{\kmax}$ that was suggestively named after its effect of jumping between valleys changing $\kmax$. One should also stress that it is a non-trivial statement that 126 configurations tend to be rather coherent -- again strengthening the association of $\tau_{2}$ with $\tau_{\kmax}$.

\subsection{Analysis and reasoning about the rest of parameter space}
The previous example indicates two important facts: Firstly, our choice of algorithmic parameters rendered $\tau_{\mathrm{fluct}}$ negligible while $\tau_{\kmax}$ is of considerable size. Secondly, besides $\tau_{\kmax}$ being large it is still \textit{under control} in the sense that we have a statistically significant number of independent configurations $N_{\mathrm{MC}}/\tau_{\kmax} \approx 380$ even in the worst case discussed above. As the chosen parameter set is well inside the region of intermediate-scale inhomogeneous order we conclude from a statistically robust ensemble that our claims of intermediate-scale inhomogeneities without spontaneous symmetry breaking of any kind are robust with respect to autocorrelation effects. We checked for similar examples on all lattice sizes and lattice spacings.

However, the above example was taken from a moderate temperature region. As we confirmed in this study, at the $T=0$ line the system is critical which implies diverging correlation lengths -- also in the MC-time evolution as is well known around practitioners \cite{Janke:2008}. We therefore expected and \emph{a posteriori} verified huge autocorrelations for temperatures close to zero. One should stress that this is a physical effect; it can likely be circumvented by an appropriately adapted algorithm but still bares physically relevant information. Still, for a small region of very low temperatures $\tau_{\kmax}$ can easily exceed our greatest efforts of up to $8\cdot10^4$ configurations generated for some parameter sets. We therefore suggest to view the very-low-temperature results with a grain of salt quantitatively: They surely give a good impression of what phenomena to expect in the exceedingly large regions of space that are correlated for these temperatures but they might be quantitatively off due to autocorrelations suppressing the interference from subdominant local minima.

We remark that $\tau_{\kmax}$ has a clear tendency to decrease in the infinite-volume limit. This is the opposite behavior of what is typically found in symmetry-breaking systems and considered further evidence in support of the existence of a BKT phase and against spontaneous symmetry breaking, as was also mentioned in the main text.

The effect of larger flavor numbers is to reduce quantum fluctuations or, in the pictorial language from above, to deepen the valleys and grow the ridges. This effect is responsible for ultimately obtaining actual spontaneous symmetry breaking in the limit of infinite number of flavors. It also greatly enhances autocorrelations, particularly in suppressing jumps between different valleys. For $\Nf=8$ the temperatures necessary to clearly observe inhomogeneities on average and jumps in the dominant wave number during the MC evolution occur concurrently are higher than in the 2-flavor case. This strengthens our finding in this and previous papers \cite{LPW20,LPW20_1} that the convergence to mean-field behavior is quite rapid with the number of flavors. While technically this casts some doubt on the quantitative accuracy of the $\Nf=8$ data, we want to point out again that this is driven by physical properties of the system more than technical difficulties.

Finally, we want to share some thoughts on how to improve on the situation: Due to extensive analytical results about the effective action, we were able to obtain a clear picture of the cause of autocorrelations in this model. One can easily imagine algorithmic improvements leveraging this knowledge. As the local minima can be enumerated by their dominant wave number $\kmax$, local updates in Fourier space might suffice, e.g.\@ local Metropolis-like updates or swapping of Fourier components. As the current updating scheme is very efficient to reduce some part of the autocorrelations, it would probably be advantageous to combine both kinds of updates into a single update step. Similar ideas are already used, e.g.\@ \cite{Albandea:2021lvl}. Another approach could be to constrain the simulations to a single sector and \emph{a posteriori} combining the results into a weighted sum.

However, these approaches would be very specific and hardly generalizable to related problems, e.g.\@ topological freezing in lattice QCD \cite{Alles:1996vn}. A modern approach that is agnostic of analytical knowledge, which is usually not as easily employed in more realistic theories, could be independence samplers from generative models, i.e.\@ independently drawing the configurations from an efficient approximation of the probability distribution. Promising results in this direction were presented in \cite{Kanwar:2020xzo}, where they overcame topological freezing in 1+1D $\mathrm{U}(1)$ gauge theory.
	\section{On fermionic 4-point functions}
	We aim at relating the U$_A$(1)-invariant fermion $4$-point function
(\ref{eq:wi1}) to the spatial correlation functions $C_{\sigma\sigma}$ and $C_{\sigma\pi}$ 
defined in \eqref{eq:spatial_correlators}.
To find such a relation we exploit the following 
Dyson-Schwinger equations, which can be derived in analogy to \eqref{eq:ward_identities}:
\begin{eqnarray}\label{apc:SDequations}
\begin{aligned}
	\langle \bar{\psi}\psi(\vx)\bar{\psi}\psi(\vy)\rangle &= -\kappa^2\langle\sigma(\vx)\sigma(\vy)\rangle + \kappa\delta^2(\vx-\vy)\;,\\
	\langle \bar{\psi}\psi(\vx)\bar{\psi}\gamma_*\psi(\vy)\rangle &= \ii\kappa^2\langle\sigma(\vx)\pi(\vy)\rangle\;,\\
	\langle \bar{\psi}\gamma_*\psi(\vx)\bar{\psi}\psi(\vy)\rangle &= \ii\kappa^2\langle\pi(\vx)\sigma(\vy)\rangle\;,\\
	\langle \bar{\psi}\gamma_*\psi(\vx)\bar{\psi}\gamma_*\psi(\vy)\rangle &= \kappa^2\langle\pi(\vx)\pi(\vy)\rangle - \kappa\delta^2(\vx-\vy)\;,
\end{aligned}
\end{eqnarray}
wherein we used the abbreviation
\begin{equation}
\kappa=\frac{\Nf}{g^2}\,.\label{apc:def}
\end{equation}
Recalling that $\Delta=\sigma+\ii\pi$ we can write the invariant $4$-point
function as
\begin{equation}
	C_\mathrm{4F}(\vx;\vy) = -\kappa^2
	\langle\Delta^*(\vx)\Delta(\vy)\rangle
	+2\kappa\,\delta^2(\vx)\;,\label{apc:wi1}
\end{equation}
and the axial U$_A$(1) symmetry implies (cf. \eqref{corr-rel})
\begin{equation}
\langle\Delta^*(\vx)\Delta(\vy)\rangle=
2\langle\sigma(\vx)\sigma(\vy)\rangle
+2\ii\langle\sigma(\vx)\pi(\vy)\rangle\;.\label{apc:wi3}
\end{equation}
In analogy to the spatial correlation functions (\ref{eq:spatial_correlators})
for the condensate fields we introduced the spatial correlation function for 
the $\Nf$ Fermi fields on the lattice in \eqref{eq:wi5}.
Inserting (\ref{apc:wi3}) into (\ref{apc:wi1}) we can relate (\ref{eq:wi5}) to 
(\ref{eq:spatial_correlators}) as follows:
\begin{equation}
C_\mathrm{4F}(x) = -2\kappa^2\left(C_{\sigma\sigma}(x)
+\ii\,C_{\sigma\pi}(x)\right)+2\kappa\delta(x)\;,\label{apc:wi5}
\end{equation}
where on the lattice the spatial $\delta$ distribution on the right-hand
side turns into the Kronecker symbol $\delta_{x,0}$.
	\label{app:4point}	
	\section{Parameters}
	\label{app:parameters}
	\begin{table}[h]
	\caption{\label{t:parameters_scale_setting} 
		Lattice spacings for each set of $(\Nf, \Ns, g^2)$. 
		The right column contains the values of
		$\Nt$ at which the scale setting was performed and
		the asterisk on one value of $a\rho_{0}$ indicates that the 
		given uncertainty was estimated by hand to be larger than 
		the computed jackknife uncertainty due to
		small statistics.}
	\newcommand{\ncol}{4}
	\renewcommand{\tabcolsep}{4pt}
	\begin{tabular}{ccccc}
		\hline \hline
		$\Nf$ & $\Ns = L / a$  &    $1/g^2$ & $a \rho_0$&$\Nt$\\
		\hline
		\rule[-1ex]{0pt}{5ex}
		$2$& $63$ & 1.0540&$0.45655 \pm 0.00061$&$72$\\
		\hline
		\rule[-1ex]{0pt}{5ex}
		$2$& $127$ & 
		1.0540&$0.45844 \pm 0.00095$&$72$\\
		\hline
		\rule[-1ex]{0pt}{5ex}
		$2$& $255$ & 
		1.0540&$0.4573 \pm 0.0012$&$72$\\
		\hline
		\rule[-1ex]{0pt}{5ex}
		$2$& $127$ & 
		1.3895&$0.1904 \pm 0.0027$&$240$\\
		\hline
		\rule[-1ex]{0pt}{5ex}
		$2$& $255$ & 
		1.8254&$0.08 \pm 0.01^{*}$&$648$\\
		\hline
		\rule[-1ex]{0pt}{5ex}
		$8$& $63$ & 
		5.1013&$0.41235 \pm 0.00023$&$80$\\
		\hline
		\hline
	\end{tabular}
\end{table}

In order to calculate the various phase diagrams we generated many ensembles 
characterized by the control parameters $(\Nf,T,L,\mu)$ or 
$(\Nf,N_t,N_s,\rho_0\mu)$, plus the four-Fermi coupling $g^2$ tuned
to the required lattice spacing measured in units of $\rho_0=\langle\rho\rangle_{T=\mu=0}$.
We summarize the lattice spacings corresponding to
the different values of $\Nf$, $\Ns$ and $g^2$ in \tref{t:parameters_scale_setting}.

As explained in the main text, we used different initial 
conditions for the fields to deal with thermalization problems: We performed
scans with Gaussian-distributed seeds with mean zero, a freeze-out from high temperatures to 
reduce thermalization times and a heat-up procedure from the 
lowest temperature to exclude any hysteresis 
effects from the freeze-out. We also used a homogeneous cold start, 
in the sense of setting the initial configuration to $\Delta({\vx})=1+\ii$ for all $\vx$, 
at small $\mu$, where inhomogeneous configurations are suppressed.
In \tref{t:control_parameters} we collect 
the control parameters $\Nt$ and $\mu$ for which we generated ensembles
in equilibrium for each of these methods. Notice that we use the same lattice spacings 
as in \tref{t:parameters_scale_setting}, which were determined via the freeze-out procedure,
irrespective of the initial conditions.

\begin{table*} 	
	\caption{\label{t:control_parameters} Parameter sets used in 
		the simulations. Note that the uncertainty of $a\rho_{0}$ 
		(from \tref{t:parameters_scale_setting}) propagates to the values 
		of $\mu/\rho_{0}$, although we did not make this explicit for the 
		sake of readability.}
	\renewcommand{\arraystretch}{1.5}
	\newcommand{\ncol}{4}
	\renewcommand{\tabcolsep}{18pt}
	\begin{tabular}{cccc}
		\hline \hline
		$\Ns = L / a$ &$1/g^2$   & $\Nt = 1 / T a$ & $\mu/\rho_0$ \\
		\hline
		\multicolumn{\ncol}{l}{
			\textbf{infinite-volume extrapolation $(\Nf=2$)}}\\
		\hline
		\rule[-1ex]{0pt}{6.5ex}
		$63$ &1.0540& 
		\shortstack{2,  4, 6, 8, 10, 12, 16,\\ 24, 32, 40, 48, 72}& 
		0, 0.0876, \dots,\ 1.3142\\
		\hline
		\rule[-1ex]{0pt}{6.5ex}
		$127$&1.0540 & 
		\shortstack{2,  4,  6,  8, 10, 12, 16,\\ 24, 36, 48, 72}& 
		0, 0.0873, \dots,\ 1.3088\\
		\hline
		\rule[-1ex]{0pt}{6.5ex}
		$255$&1.0540 & 
		\shortstack{2,  4,  6,  8, 10, 12, 16,\\ 24, 36, 48, 72}& 
		0, 0.0875, \dots,\ 1.3121\\
		\hline
		\multicolumn{\ncol}{l}{
			\textbf{continuum extrapolation $(\Nf=2$)}}
		\\
		\hline
		\rule[-1ex]{0pt}{6.5ex}
		$127$&1.3895 & 
		\shortstack{4,   6,   8,  10,  12,  16, 24,\\  36,  48,  72,  96, 144}& 
		0, 0.1050, \dots,\ 1.5756\\
		\hline
		\rule[-1ex]{0pt}{6.5ex}
		$255$&1.8254 & 
		\shortstack{8,  12,  16,  24,  36,\\  48,  72,  96, 144}& 
		0, 0.1250, \dots,\ 1.8750\\
		\hline
		\multicolumn{\ncol}{l}{
		\textbf{independent initial conditions ($\Nf=2$, for crosschecks)}}
		\\
		\hline
		\rule[-1ex]{0pt}{6.5ex}
		$63$ &1.0540& 
		\shortstack{16, 24, 32, 40,\\ 48, 56, 64, 80}& 
		0, 0.0876, \dots,\ 1.3142\\
		\hline
		\rule[-1ex]{0pt}{6.5ex}
		$127$&1.0540 & 
		24, 32, 40, 48, 64, 80& 
		0, 0.0873, \dots,\ 1.3088\\
		\hline
		\rule[-1ex]{0pt}{6.5ex}
		$255$ &1.0540& 
		\shortstack{8, 16, 24, 32, 40,\\ 48, 64, 80}& 
		0, 0.0875, \dots,\ 1.3121\\
		\hline
		\rule[-1ex]{0pt}{6.5ex}
		$127$ &1.3895& 
		8, 16, 40, 48, 64, 80& 
		0, 0.1050, \dots,\ 1.5756\\
		\hline
		\rule[-1ex]{0pt}{6.5ex}
		$255$ &1.8254& 
		24, 32, 40, 48, 64& 
		0, 0.1250, \dots,\ 1.8750\\
		\hline
		\multicolumn{\ncol}{l}{
			\textbf{independent initial conditions
		($\Nf=8$)}}
		\\
		\hline
		\rule[-1ex]{0pt}{6.5ex}
		$63$ &5.1013& 
		\shortstack{4, 6, 8, 12, 16, 20, 24, 28,\\32, 36, 40, 48, 56, 64, 80}& 
		0, 0.0970, \dots,\ 1.4551\\
		\hline
		\multicolumn{\ncol}{l}{
			\textbf{heat-up initial conditions ($\Nf=2$, for checking hysteresis effects)}}
		\\
		\hline
		\rule[-1ex]{0pt}{6.5ex}
		127& 1.0540&16, 24, 36, 48, 72& 0.4363, 0.8725, 1.3088 \\
		\hline
		\multicolumn{\ncol}{l}{
			\textbf{cold start ($\Nf=2$, small $\mu$)}}
		\\
		\hline
		\rule[-1ex]{0pt}{6.5ex}
		63&1.0540&\shortstack{2, 4, 6, 8, 10,\\ 12, 24, 48, 72}&0, 0.0876, 0.1752\\
		\hline
		\hline
	\end{tabular}
\end{table*}

\FloatBarrier
	
	\newpage
	\bibliography{bibliography}

\end{document}